\documentclass[letterpaper]{article} % DO NOT CHANGE THIS
\usepackage{aaai25}  % DO NOT CHANGE THIS
\usepackage{times}  % DO NOT CHANGE THIS
\usepackage{helvet}  % DO NOT CHANGE THIS
\usepackage{courier}  % DO NOT CHANGE THIS
\usepackage[hyphens]{url}  % DO NOT CHANGE THIS
\usepackage{graphicx} % DO NOT CHANGE THIS
\usepackage{natbib}  % DO NOT CHANGE THIS AND DO NOT ADD ANY OPTIONS TO IT
\usepackage{caption} % DO NOT CHANGE THIS AND DO NOT ADD ANY OPTIONS TO IT
\usepackage{algorithm}
\usepackage{algorithmic}
\usepackage{booktabs}
\usepackage{multirow}
\usepackage{multicol}
\usepackage{makecell}
\usepackage{lipsum}
\usepackage{array}
\usepackage{amsmath}
\usepackage{amssymb}

\usepackage[dvipsnames]{xcolor}
\newcommand{\answerYes}[1]{\textcolor{blue}{#1}} 
\newcommand{\answerNo}[1]{\textcolor{teal}{#1}} 
\newcommand{\answerNA}[1]{\textcolor{gray}{#1}}

\usepackage{newfloat}
\usepackage{listings}
\DeclareCaptionStyle{ruled}{labelfont=normalfont,labelsep=colon,strut=off} % DO NOT CHANGE THIS
\floatstyle{ruled}
\newfloat{listing}{tb}{lst}{}
\floatname{listing}{Listing}
\title{How Growing Toxicity Manifests: A Topic Trajectory Analysis of U.S. Immigration Discourse on Social Media}

\author {
    Una Joh\textsuperscript{\rm 1},
    Yiqi Li\textsuperscript{\rm 1},
    Jeff Hemsley\textsuperscript{\rm 1}
}
\affiliations {
    \textsuperscript{\rm 1}School of Information Studies, Syracuse University\\
    sjoh01@syr.edu, yli360@syr.edu, jjhemsle@syr.edu
}

\begin{document}

\maketitle

\begin{abstract}
In the online public sphere, discussions about immigration often become increasingly fractious, marked by toxic language and polarization. Drawing on 4 million X posts over six months, we combine a user- and topic-centric approach to study how shifts in toxicity manifest as topical shifts. Our topic discovery method, which leverages instruction-based embeddings and recursive HDBSCAN, uncovers 157 fine-grained subtopics within the U.S. immigration discourse. We focus on users in four groups: (1) those with increasing toxicity, (2) those with decreasing toxicity, and two reference groups with no significant toxicity trend but matched toxicity levels. Treating each posting history as a trajectory through a five-dimensional topic space, we compare average group trajectories using permutational MANOVA. Our findings show that users with increasing toxicity drift toward alarmist, fear-based frames, whereas those with decreasing toxicity pivot toward legal and policy-focused themes. Both patterns diverge statistically significantly from their reference groups. This pipeline, which combines hierarchical topic discovery with trajectory analysis, offers a replicable method for studying dynamic conversations around social issues at scale. \footnote{This is the preprint of a paper accepted at ICWSM 2026. The copyrighted version will be available at the AAAI Digital Library (https://ojs.aaai.org/index.php/ICWSM) once published.}
\end{abstract}

% Uncomment the following to link to your code, datasets, an extended version or similar.
%
% \begin{links}
%     \link{Code}{https://aaai.org/example/code}
%     \link{Datasets}{https://aaai.org/example/datasets}
%     \link{Extended version}{https://aaai.org/example/extended-version}
% \end{links}

\section{Introduction}
Social media is an increasingly indispensable public sphere, offering a democratized space for diverse discussion. However, alongside the problems such as fragmentation, addiction, misinformation, toxicity is another critical issue that may emerge on social media, potentially leading to harms such as opinion distortion, biases, hostility, polarization, and even real-world violence \cite{bruns_is_2015, kim_distorting_2021, gallacher_online_2021, klein_medium_2024}. Immigration, one of the most contentious topics discussed everyday on social media, exemplifies the diversities and complexities of social media discussion \cite{mittos_analyzing_2020}. For example, as per Pew Research’s recent report, immigration was a major point of contention in the 2024 presidential election, deepening the divide between Trump and Harris supporters \cite{mukherjee2024trump}. Immigration has been raised by existing scholarship to be a space contaminated by toxic content and divisive debates \cite{santana_incivility_2015}.  This study draws from a rich, longitudinal, and comprehensive immigration-related social media data, and explores the dynamics of toxicity in the online public sphere of X (formerly known as Twitter). 

Conceptually, this research is novel in a few ways. Firstly, it bridges the isolation between social media toxicity research that are centered on topics (e.g., \cite{klein_medium_2024,rossini_characterizing_2021,stromer-galley_context_2015}), and those that revolve around toxic users or behaviors (e.g., \cite{rajadesingan_quick_2020,coe_online_2014}). We found that users with increasing toxicity and those with decreasing toxicity engage with distinctive sets of topics, differing from the topic trajectories of users with same average toxicity level with each groups. This highlights that toxicity in communication can follow complex and varied topical pathways rather than simply representing a static trait of individual users.

Secondly, under the backdrop of this rich and contentious immigration issue public sphere, seldom is there longitudinal observations on the co-evolutionary patterns of users' dynamic engagement across topics, and toxicity communication patterns. Lastly, while most existing research studies extreme toxic behaviors or toxic accounts \cite{qayyum2023longitudinal,kumar2022understanding}, this current research sheds light on users of different toxic tendencies (specifically, escalating or reducing toxicity over time). Such a focus not only provides unique angle on a group of shifting and impacted individuals in the issue community, but also drives inquiries of associated factors linked with toxicity changes. For example, we found that as users’ toxicity rises, they gravitate toward alarmist, threat-framed narratives, whereas users whose toxicity declines increasingly engage with procedural or policy-oriented themes.

In addition, this study makes methodological contributions. We refine topic discovery models by incorporating instruction-based embedding and a recursive HDBSCAN clustering framework with hierarchical merging based on topic coherence. This technique mitigates the challenge of over-segmentation, yielding interpretable subtopics that capture both the semantic and stance-related nuances of immigration discourse. Moreover, we pair this topic-discovery framework with a trajectory-analysis pipeline that models each user’s posting history as a continuous path through embedding space and statistically compares group trajectories, offering a scalable way to link topical movement with shifts in toxic behavior.

\section{Related Work \& Research Questions}
\subsection{Toxicity on Social Media}

Social media platforms like X reflect Habermas' concept of the public sphere, increasingly playing a critical role in facilitating democratized, open, real-time, and diverse discussions. However, the social media public sphere is also highly fragmented, characterized by toxic conversations, biased viewpoints, and increasing polarization \cite{bruns_is_2015}. Catalyzed by factors such as the social media algorithms, social influences, and emotional contagion, toxicity on these platforms is significantly impacting participants, distorting public opinion, fueling negativity, and exacerbating polarization \cite {kim_distorting_2021}. Social media toxicity harms public discourse by triggering irrational discussions, obstructing productive conversation, and engendering shallow deliberation \cite{klein_medium_2024}. As the emotional foundation of toxic language further spreads through social media networks, adverse outcomes may also emerge, including the spread of negativity, hostility, polarization, or even real-world violence across groups \cite{gallacher_online_2021, klein_medium_2024}. 

Toxicity is defined as communication ``manifest(ed) in the tone and style with which a speaker attacks their addressee’s `face,' or public self-image'' \cite[p. 5]{sydnor_signaling_2019}. A plethora of social media toxicity research falls into topic-driven or user-centered realms. Topic-wise, toxicity is often linked with contextual factors such as information sources or topics \cite{klein_medium_2024}. Existing research identifies that ``hard news'' tends to generate high incivility or toxicity, while lighter topics such as lifestyle or technology are linked with reduced toxicity \cite{klein_medium_2024}. When theme of the toxic content is targeting different groups (especially LGBTQ population), audiences are likely to seek for content moderation from the platform, although the general motivation of moderation-seeking is limited \cite{pradel_toxic_2024}. Political topics are also especially likely to provoke toxic discussion \cite{chen_misleading_2022, rossini_characterizing_2021, stromer-galley_context_2015}. Immigration is among a few topics (e.g., climate change, genetic testing) that are especially toxicity-prone \cite{mittos_analyzing_2020, salminen_topic-driven_2020, santana_incivility_2015}. 

From a user-centered perspective, researchers mainly study what user-level factors contribute to toxic communication and community behaviors. For example, Rajadesingan and colleagues found that pre-entry learning allows newcomers into the Reddit communities to conform to the community's preexisting toxicity norms \cite{rajadesingan_quick_2020}. Coe and colleagues found that frequent commenters on newspaper websites tend to be less toxic in comments than less frequent commenters \cite{coe_online_2014}. Importantly, many studies focus on extreme behaviors by tracking toxic profiles \cite{qayyum2023longitudinal,kumar2022understanding}, this study takes a novel approach of identifying types of invested discussants from a longitudinal angle, exploring users who are naturally engaged in an issue discussion, and examining over-time engagement patterns—specifically those whose toxicity increases and decreases within an issue space \cite{yang_issue_2020}. 

The topic-centered and user-centered perspectives of toxicity are often approached in isolation. Such isolation risks oversimplifying toxicity by missing the potential interplay between user behaviors and topics. To address this gap, this study examines longitudinal patterns of users' over-time toxic communication patterns, and explore the interplay between different types of users and their topic-engagement. This way, we are able to map out how the topical and user-level toxicity level co-evolve, thus providing valuable empirical observations on where the toxicity may emerge, and where cross-group dialogue and shared values that reduce toxicity may develop. 

This study situates against the backdrop of longitudinal immigration-related discussion on X. As discussed, immigration is one of the most polarized and emotionally charged issues \cite{mittos_analyzing_2020, salminen_topic-driven_2020}, strongly associated with many related issues of discussion such as racial and ethnicity, unemployment, crime, border safety, economic outlook, social justice, and more \cite{santana_incivility_2015}. Empirical analysis on the discourse around immigration informs understanding of how the changing and controversial policies (e.g., DACA, refugee cities, border security) are linked with audiences' evolving communication and engagement patterns. Immigration can be conceptualized as a social issue space and a bounded issue ecology because it ``channels public attention and provides a space for the communication of identities and ideologies'' \cite[p. 9]{yang_issue_2020}.

\subsection{Research Questions}
As outlined in the previous subsection, research on social media toxicity typically adopts one of two primary perspectives: topic-centric or user-centric. While each approach has yielded valuable insights, they are often treated in isolation.  To address this gap, this study integrates both perspectives by identifying user groups with increasing or decreasing toxicity levels and analyzing the trajectories of the subtopics these groups engage with. We statistically compare these groups to corresponding reference groups (i.e., users with the same average toxicity as each group, but without a statistically significant trend of increasing or decreasing toxicity). This comparison aims to determine whether differences in subtopic trajectories are associated with changes in toxicity over time, independent of overall toxicity levels, thereby uncovering the thematic contexts that accompany escalation or de-escalation. Accordingly, the research questions of this study are:

\begin{enumerate}
    \item \textbf{RQ1 (Increasing Toxicity Users):} \\
    Are the temporal topic trajectories of users with significant toxicity \textit{increases} different from those of a toxicity-matched reference group without such increases?

    \item \textbf{RQ2 (Decreasing Toxicity Users):} \\
    Are the temporal topic trajectories of users with significant toxicity \textit{decreases} different from those of a toxicity-matched reference group without such decreases?
\end{enumerate}

\section{Data}
To answer our research questions, we collected data from X using U.S. immigration-related keywords. X was chosen because it is one of the major social media platforms in the U.S. context \cite{gottfried_americans_2024} and is also known for its abundance of toxicity, especially in recent times \cite{hickey_auditing_2023}. Due to the limitations of X's official API, we used Apify, which is a web scraping and automation platform, to collect the data.

The collection time frame was from April 17, 2023, to October 27, 2023. This timeframe was selected because, according to a poll \cite{jones_sharply_2024}, this period coincided with a noticeable worsening of sentiment toward immigration issues. We concluded that using this timeframe would allow us to capture a substantial portion of the dynamics of user toxicity change.

Data collection was performed using keywords. According to the API provider, they extracted posts from the web search feature of X. This feature offered two options, ``latest'' and ``top,'' but the distinction between these options was not clearly explained in X's official documentation. After consulting with the API provider, we chose the ``latest'' option to collect the most comprehensive data, following their advice. Additionally, we restricted the language option to English because it was an economical way to filter out a substantial number of posts regarding non-U.S. immigration issues, despite the potential limitation this poses for the scope of discourse we can collect.

The search query used for the collection was as follows:
\begin{quote}
\textit{(immigrant OR immigrants OR immigration OR migrant OR migrants OR migration OR illegals OR undocumented OR refugee OR refugees OR ``guest worker'' OR ``guest workers'' OR ``asylum seeker'' OR ``asylum seekers'' OR ``illegal alien'' OR ``illegal aliens'') AND (USA OR ``U.S.'' OR ``United States'' OR ``the US'' OR America OR American OR Americans OR Biden OR Trump)}
\end{quote}

We opted for broad terms rather than attempting to construct an exhaustive list of keywords because X's web UI does not function properly when too many keywords are provided as input. Additionally, X's documentation does not clearly specify the maximum number of keywords allowed.

In total, we collected 8,995,234 posts, including original posts, quotes, and replies. The API did not have a feature to collect reposts (retweets), which was less important for our research since we aimed to trace changes in the toxicity of individual users' posts.

\section{Methodology}
\subsection{Classification for Filtering Relevant Posts}
Although the search query was designed to retrieve relevant posts about U.S. immigration issues, it was inevitable to collect some irrelevant posts due to the limitations of keyword-based retrieval. To address this, we filtered out irrelevant posts from the dataset.

Given the strong performance of decoder-based large language models (LLMs) in classification tasks within social science contexts \cite{ziems_can_2024}, we employed a decoder-based LLM to filter out irrelevant posts. Considering the simplicity of the task, we performed zero-shot classification and evaluated the results using the F1 score.

Each post was processed independently as input, with the classification task divided into two steps. The prompt for Task 1 was: ``Is this tweet about immigration? Answer with either `Yes' or `No.' ''  Only posts classified as ``Yes'' in Task 1 proceeded to Task 2, which used the following prompt: ``Is this tweet in a U.S. context? Answer with either `Yes' or `No.' ''

Due to the simplicity of the task, we used a small-sized LLM, the unquantized Gemma2-9B-instruct\footnote{https://huggingface.co/google/gemma-2-9b-it} model, which has a context window size of 8,192 tokens. This was sufficient for our use case, as each input consisted of only one question and one post. To ensure reproducibility, we set the temperature to 0 and the top-p value to 0.9. The model was accessed via the Deep Infra\footnote{https://deepinfra.com} API, with a total processing cost of \$14.66.

To validate the classification performance, we conducted human coding on 300 posts prior to running the zero-shot classification with the Gemma2 model. The human coders included one of the authors and a fourth-year Ph.D. student at a U.S. institution. Inter-coder reliability was measured using Cohen's Kappa \cite{cohen_coefficient_1960}, yielding scores of 0.86 for Task 1 and 0.79 for Task 2. Discrepancies were resolved through discussions among the coders to establish the gold label.

Gemma 2 achieved an F1 score of 0.92 on Task 1 and 0.81 on Task 2 when evaluated against the gold labels. Most errors were false positives, indicating that the model is more permissive than the human annotators.
 In Task 1 (immigration vs. other), for example, Gemma 2 tagged a post reminiscing about photographing the annual migration of birds as “immigration‐related,” presumably because of the lexical overlap with migration. Likewise, a discussion of Canadian elections that mentioned the author’s immigrant background was flagged as immigration content even though human immigration policy was not the focus. Typical misclassifications in Task 2 (U.S. vs. non-U.S.) include a post criticizing government spending on illegal immigrants in hotels that Gemma2 labeled as U.S.-specific despite wording that is also common in the U.K.

Fortunately, the issue of false positives is less of a problem than false negatives, as those false positives are likely mostly filtered out during the clustering process as outliers or small clusters. After filtering, a total of 4,651,275 posts were used for the subsequent analysis.

\subsection{Toxicity Assessment}
To assess the toxicity of posts on X, we once again leveraged large language models (LLMs) instead of using Google's Perspective API, which is the most prevalent tool for evaluating toxicity in user-generated online content \cite{gervais_incivility_2025}. The primary reason for not using the Perspective API is that its outputs represent the likelihood of toxicity, rather than providing a true measure of severity, as \citeauthor{gervais_incivility_2025} (2025) has noted. Another widely used alternative, the Detoxify library \cite{Detoxify}, shares this limitation, as it is trained on the same dataset as the Perspective API.

We explored the potential of LLMs to offer an alternative approach by prompting the models to evaluate toxicity in a more human-like manner. In a recent study, \citeauthor{de_wynter_rtp-lx_2025} (2025) investigated whether LLMs can serve as reliable toxicity evaluators across multiple languages, introducing RTP-LX, a human-annotated benchmark. Their findings showed that toxicity ratings generated by GPT-4 Turbo aligned closely with human annotations on the English subset of the RTP-LX dataset.

In our study, we experimented with two different LLMs to rate toxicity on a Likert-type scale ranging from 1 (benign) to 5 (highly toxic): OpenAI’s GPT-4.1-nano-2025-04-14 and Google’s Gemma3-4B-Instruct\footnote{\url{https://huggingface.co/google/gemma-3-4b-it}}. Given the size of our dataset, we limited our evaluation to smaller models to reduce computational costs. Both models were configured with a temperature of 0 and a top-p value of 0.9.

For each text sample in the English subset of RTP-LX, we prompted the models to output a toxicity score using a slightly modified version of the prompt provided in the original RTP-LX paper \cite{de_wynter_rtp-lx_2025}, tailored to our context (see Appendix B). We then compared the LLM-generated scores to ground-truth human ratings. The OpenAI model achieved a Pearson correlation of 0.7701 with the human ratings, while the Gemma-3B model yielded a correlation of 0.7204. Our results suggest that LLMs can approximate human judgments of toxicity with a reasonably strong degree of correlation, especially considering the inherently subjective nature of the task.

Despite the superior performance of GPT-4.1-nano, we opted for the Gemma3-4B-Instruct model via the Deep Infra API, primarily for cost-efficiency, given that the OpenAI API was nearly ten times more expensive. The total cost using the Gemma model was \$20.52. For our subsequent analysis, we normalized toxicity scores from the 1–5 Likert scale to a 0–100 scale.

\subsection{Topic Discovery}
\label{sec:topic-discovery}

Identifying subtopics in U.S. immigration-related discourse is a crucial task for our study. Since latent Dirichlet allocation (LDA) was first introduced by Blei et al. \shortcite{blei_latent_2003}, numerous topic modeling techniques have been proposed \cite{vayansky_review_2020}. In particular, with the advancement of natural language processing and deep neural networks, a variety of topic models have been developed, broadly categorized as neural topic models \cite{wu_survey_2024}.

A characteristic of these models is their probabilistic nature. Specifically, probabilistic topic models assume that topics are defined as probability distributions over keywords. Documents are generated as sequences of words sampled from these topics. The sampling follows the probability distribution of topics assigned to the document and the probability distribution of keywords within each topic.

Criticism of these assumptions in probabilistic topic models has led to the exploration of simpler frameworks, such as embedding-clustering-based topic discovery models \cite{thompson_topic_2020, angelov_top2vec_2020, zhang_is_2022}. These models avoid the assumption that documents are generated based on probabilistic distributions, often categorized as ``topic discovery'' models rather than traditional ``topic modeling'' approaches \cite{wu_survey_2024}. One such topic discovery model that has gained popularity among social science researchers is BERTopic \cite{grootendorst_bertopic_2022}. BERTopic combines document embedding using Sentence-BERT \cite{reimers_sentence-bert_2019} with clustering using HDBSCAN \cite{campello_density-based_2013}.

Our topic discovery method builds upon BERTopic but differs in two significant ways: 
\begin{enumerate}
    \item Instruction-based document embedding, and
    \item Hierarchical topic merging based on topic coherence.
\end{enumerate}
These modifications are designed to address specific limitations in the original BERTopic approach and better align with the needs of our study.

\subsubsection{Instruction-based Document Embedding}

In this work, we leverage an instruction-based document embedding model, first introduced by INSTRUCTOR \cite{su_one_2023}. Unlike S-BERT \cite{reimers_sentence-bert_2019} and SimCSE \cite{gao_simcse_2021}, which were designed for generating general-purpose document embeddings, instruction-based embedding models allow more task-specific embeddings by incorporating instructions with the input text \cite{su_one_2023}.

While general-purpose embedding models are useful for many applications, they have limitations in distinguishing nuanced differences in semantic meaning, especially for tasks requiring an understanding of context or stance. For example, as noted by Introne \shortcite{introne_measuring_2023}, general-purpose embedding models often yield high cosine similarity for semantically opposite sentences, such as ``Illegal immigrants are causing problems'' and ``Illegal immigrants are not causing problems.'' This limitation arises because these models are not explicitly optimized to capture task-specific distinctions, such as differences in sentiment or stance within a given topic.

Instruction-based document embedding models address this issue by jointly taking the input text and an instruction describing the downstream task \cite{su_one_2023}. This approach enables the model to produce embeddings that align better with the task's requirements.

To investigate this, we experimented with several sentence pairs to evaluate the performance of different models. One example pair was ``Illegal immigration helps the U.S. economy by filling jobs and contributing to growth'' and ``Illegal immigration hurts the U.S. economy by taking jobs and draining resources.'' The all-mpnet-base-v2 model\footnote{huggingface.co/sentence-transformers/all-mpnet-base-v2}, which was one of the embedding models used in the original BERTopic paper, produced a cosine similarity score of 0.864 for these two sentences. While these sentences are similar in terms of topic (illegal immigration), they express opposing stances, highlighting the inability of task-agnostic embedding models to capture differences in stance.

In contrast, an instruction-based embedding model, NV-Embed-v2\footnote{https://huggingface.co/nvidia/NV-Embed-v2}, yielded more nuanced results when provided with different instructions. For instance, when instructed with ``What topic is this tweet addressing?'', the model produced a cosine similarity score of 95.90 for the two example sentences. Meanwhile, when instructed with ``What is this tweet's view on illegal immigration?'', the cosine similarity dropped to 71.87. This demonstrates that instruction-based embedding models can effectively distinguish between subtopics and stance when guided by appropriate instructions.

To simultaneously capture both the subtopics of discourse on U.S. illegal immigration and the stance on those subtopics, we constructed the following instruction for embedding: 
``This is one of the tweets about U.S. immigration issues. What is this user's stance on immigration, and which specific subtopic of immigration does this tweet address?'' We chose the NV-Embed-v2 model due to its superior performance on the MTEB benchmark \cite{muennighoff_mteb_2023} as of November 28, 2024, according to the Hugging Face leaderboard.\footnote{https://huggingface.co/leaderboard} The embedding process took a total of 107 hours and 28 minutes using an NVIDIA RTX A6000 GPU.

\subsubsection{Recursive Clustering}

To cluster the embedded posts, dimensionality reduction was necessary to mitigate the curse of dimensionality \cite{aggarwal_surprising_2001}. Following prior literature \cite{grootendorst_bertopic_2022}, we reduced the dimensionality of the embeddings from 4,096 to 5 dimensions using UMAP \cite{mcinnes_umap_2020}. UMAP was chosen because it performs well in preserving the global and local structure of data compared to alternatives like t-SNE \cite{maaten_visualizing_2008} or PCA \cite{mackiewicz_principal_1993}, while also being computationally efficient.

Given our dataset of over 4 million vectors, each with 4096 dimensions, it was not computationally feasible to apply UMAP to the entire dataset. Instead, we randomly sampled 10\% of the dataset, as this subset was sufficient to capture the global structure of the data and train the UMAP model effectively. The trained model was then used to project the remaining 90\% of the dataset, reducing all embeddings to 5-dimensional vectors.

As the final step, we employed the HDBSCAN \cite{campello_density-based_2013} algorithm for clustering. HDBSCAN, a density-based clustering algorithm, was selected for its ability to discover clusters of arbitrary shapes and handle noise effectively by not forcing all data points into clusters. Additionally, it offers advantages over DBSCAN \cite{ester_density-based_1996} by handling clusters with varying densities through its hierarchical clustering approach.

We experimented with six parameter sets for \texttt{min\_cluster\_size} and \texttt{min\_samples}: (100, 200, 300, 1000, 2000, 3000). Parameters of 100 and 200 resulted in over 500 clusters, introducing excessive granularity. A value of 300 caused memory issues, leading to time-out errors. Thus, we proceeded with the parameter sets of 1000, 2000, and 3000. The Euclidean distance metric was used for clustering, as the reduced 5-dimensional space was compact enough for this metric to perform effectively.

Despite these optimizations, initial results from a single pass of HDBSCAN revealed a significant imbalance in clustering. Approximately 90\% of the data was assigned to a single dominant cluster, with only a few small clusters capturing the remaining points. This imbalance likely stemmed from HDBSCAN's method of constructing density hierarchies and selecting clusters that are most stable across a range of density thresholds. When a dominant cluster spans a broad range of densities, it tends to absorb points that could otherwise form subclusters, particularly when low-density regions act as bridges between subclusters.

To address this issue, we adopted an iterative clustering strategy. In this approach, we identified all clusters larger than the \texttt{min\_cluster\_size} parameter and ran additional passes of HDBSCAN on each large cluster independently. This recursive clustering allowed the algorithm to focus on narrower density ranges within each large cluster, revealing subclusters that were initially masked. While HDBSCAN is inherently a hierarchical clustering algorithm and its tree structure theoretically captures all possible clusters, it was impractical to directly use the full hierarchy due to the overwhelming number of transient clusters across millions of data points.

\subsubsection{Hierarchical Topic Merging Based on Topic Coherence}

Using the three sets of parameters (\texttt{min\_cluster\_size} = \texttt{min\_samples} = 1000, 2000, 3000), we obtained different numbers of unique clusters. The maximum subcluster levels for each hyperparameter set were 5, 6, and 5, respectively.

Retaining all subclusters could lead to over-segmentation, as some clusters may arise from minor density fluctuations or noise within the parent cluster. To address this, we aimed to retain only those subclusters that demonstrated statistically significantly higher coherence compared to their parent cluster.

Although several methods exist for evaluating the coherence of topics generated by a topic modeling algorithm, most rely on keywords, which are not directly applicable to our approach. Instead, we adopted a qualitative evaluation method inspired by Newman et al. \shortcite{newman_external_2009}, wherein human annotators rate the quality of topics based on their coherence. To scale this method, we used the unquantized Llama-3.3-70B-instruct model\footnote{https://huggingface.co/meta-llama/Llama-3.3-70B-Instruct} to evaluate topic coherence automatically.

For the evaluation of each subcluster, we randomly sampled 30 posts from within the subcluster and 30 posts from outside it. These 60 sample posts were provided to Llama-3.3 using a standardized prompt (details in Appendix A). The model returned a single integer score on a 5-point Likert scale, representing the coherence of the topic. Symbolically, this classification can be represented as:
\[
\mathrm{Coherence} = f\bigl(x_1, x_2, \ldots, x_{30}, \; y_1, \ldots, y_{30}\bigr),
\]

where \(x_n\) are in-cluster samples, \(y_n\) are out-of-cluster samples, and \(f\) represents the classifier constructed by the language model and prompt. To improve statistical reliability, we repeated this sampling and scoring process 30 times for each subcluster, generating a distribution of coherence scores. On average, the 95\% confidence interval for the coherence scores across all subclusters was $\pm 0.08$, with the smallest CI being $\pm 0.00$ and the largest $\pm 0.62$.

Once we obtained coherence scores for each subcluster and its parent cluster, we compared the distributions using the Mann-Whitney U test \cite{nachar_mann-whitney_2008}. This non-parametric test evaluates differences in median values and allowed us to determine whether the subcluster's coherence was significantly higher than its parent cluster. Subclusters that failed to demonstrate significant improvement or had significantly lower coherence scores were merged back into their parent cluster.

Table~\ref{tab:after-merging} summarizes the clustering results after subcluster merging for different minimum cluster sizes. 

\begin{table}[htbp]
    \centering
    \begin{tabular}{lccc}
    \toprule
    \textbf{Minimum Cluster Size} & \textbf{1000} & \textbf{2000} & \textbf{3000} \\
    \midrule
    Level 1 Clusters & 11 & 9 & 9 \\
    Level 2 Clusters & 157 & 4 & 1 \\
    Level 3 Clusters & 9 & -- & -- \\
    \bottomrule
    \end{tabular}
    \caption{Cluster Counts by Level After Subcluster Merging for Different Minimum Cluster Sizes}
    \label{tab:after-merging}
\end{table}

After qualitative probing of the final clusters, we found that using a minimum cluster size of 1000 resulted in slight over-segmentation. However, this was acceptable compared to the results for 2000 and 3000, which produced overly coarse clusters. Therefore, we proceeded with the results from the \texttt{min\_cluster\_size} of 1000.

Finally, using the Llama-3.3-70B model again, we generated labels for each topic. Labels consisted of a 3–7 word noun phrase summarizing the topic and an approximately 100-word concise description, derived from 30 in-cluster and 30 out-of-cluster sample posts. A comprehensive list of topic labels, full topic descriptions, post counts, and toxicity levels can be found on this interactive visualization page\footnote{The page takes approximately 5 seconds to fully load.} (https://topic-immigration.onrender.com). Appendix C provides the brief topic descriptions instead of full descriptions, also generated by Llama-3.3-70B.

\subsection{Topic Trajectory Analysis}

\begin{figure}[t]
\centering
\includegraphics[width=0.9\columnwidth]{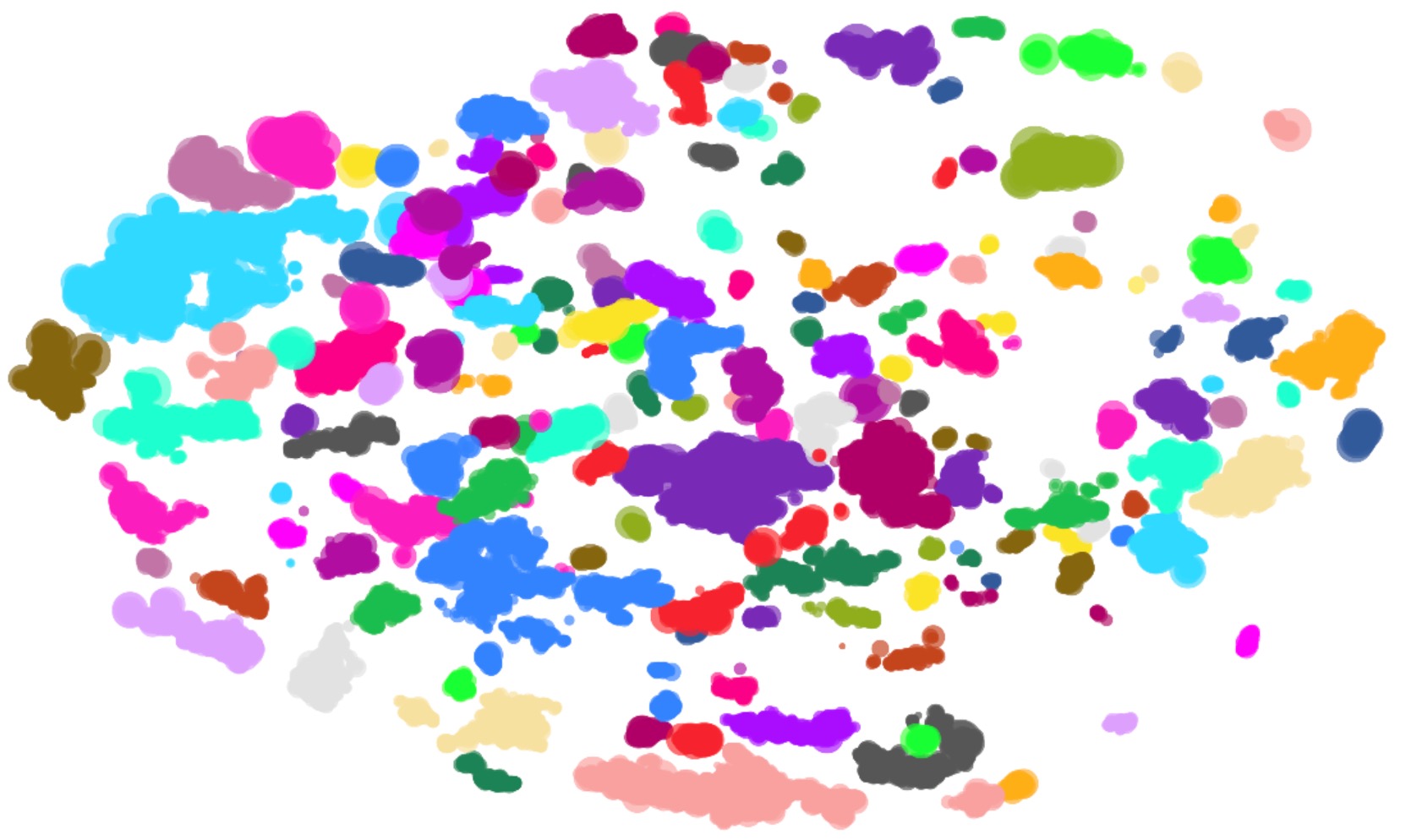} % Reduce the figure size so that it is slightly narrower than the column. Don't use precise values for figure width.This setup will avoid overfull boxes
\caption{Two-Dimensional Projection of Post Embeddings Belonging to 157 U.S. Level 2 Immigration Subtopics (Color-Coded by Topic)}
\label{viz}
\end{figure}

Based on the discovered topics, we treated them as semantically meaningful regions in the embedding space (as illustrated in Figure~\ref{viz}), where users “visit” topics by posting content associated with them over time. In this sense, each user's visiting history in topic space can be conceptualized as a \textit{topic trajectory}. 

We focus our analysis on a set of 157 topics identified as Level 2 clusters, which offer a reasonable balance between comprehensiveness and granularity. Although this level of clustering is not perfect, as will be discussed in the Results section, it provides a useful resolution for evaluating which regions of topic space users' trajectories are most closely associated with.

\subsubsection{User Grouping}
We began by identifying \emph{active users} as those who posted at least 50 times on X about U.S.\ immigration issues within our six-month study window. This yielded 8,180 active users (just 0.68\% of the 1,206,512 unique users in our dataset) who together produced 17.86\% of all US immigration-related posts.

For each active user \(i\), we modeled their toxicity score over time by fitting a simple linear regression:
\[
\mathrm{toxicity}_{it}
= \beta_{0,i} + \beta_{1,i}\,\bigl(\mathrm{timestamp}_{it}\bigr)
+ \varepsilon_{it},
\]
where \(\mathrm{toxicity}_{it}\) is the toxicity of user \(i\)’s \(t\)th post, \(\mathrm{timestamp}_{it}\) is the post time in seconds, \(\beta_{1,i}\) captures the rate of change in toxicity for user \(i\), and \(\varepsilon_{it}\) is the residual error.  We extracted those users for whom the time coefficient \(\beta_{1,i}\) was statistically significant (\(p < 0.05\)), yielding 1,124 users with a clear temporal trend in toxicity. Of these, 718 had \(\beta_{1,i} > 0\) (the \emph{Increasing Toxicity Group}) and 406 had \(\beta_{1,i} < 0\) (the \emph{Decreasing Toxicity Group}).

\begin{figure}[t]
\centering
\includegraphics[width=0.98\columnwidth]{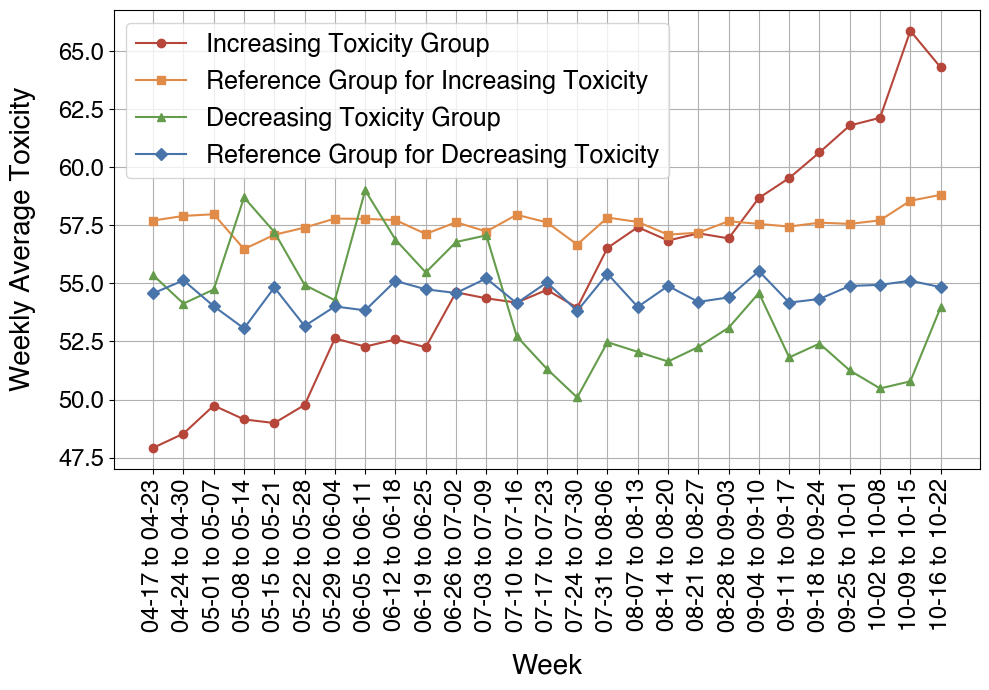} % Reduce the figure size so that it is slightly narrower than the column. Don't use precise values for figure width.This setup will avoid overfull boxes
\caption{Weekly Average Toxicity of Increasing and Decreasing Toxicity Groups and Their Reference Groups}
\label{group_tox}
\end{figure}

To isolate trajectory effects of increasing or decreasing toxicity from baseline toxicity level effects, we created matched \emph{reference groups}. First, we computed each user’s average post toxicity, $\bar{T}_i$.
The mean of these averages was \(\bar{T}_{\mathrm{inc}} = 0.5762\) for the Increasing Toxicity Group and \(\bar{T}_{\mathrm{dec}} = 0.5450\) for the Decreasing Toxicity Group. Then, among the 7,056 users without a significant trend, we selected:

\begin{itemize}
  \item 718 users whose \(\bar{T}_i\) was closest to \(\bar{T}_{\mathrm{inc}}\), forming the \emph{Reference Group for the Increasing Toxicity Group}, and
  \item 406 users whose \(\bar{T}_i\) was closest to \(\bar{T}_{\mathrm{dec}}\), forming the \emph{Reference Group for the Decreasing Toxicity Group}.
\end{itemize}
This yields four groups—(1) Increasing Toxicity Group, (2) Reference Group for the Increasing Toxicity Group, (3) Decreasing Toxicity Group, and (4) Reference Group for the Decreasing Toxicity Group—allowing us to test whether the topic trajectories of the Increasing and Decreasing Toxicity Groups differ significantly from those of their respective reference groups. Figure~\ref{group_tox} shows each group's weekly average toxicity.

\subsubsection{Linear Interpolation of Topic Trajectories}

Each user’s trajectory is represented as a temporal sequence of 5-dimensional embedding vectors derived from their posts. Because post timestamps are recorded to the nearest second, we first align all trajectories to a common 194-day grid (the full span of the study) by linearly interpolating between successive post embeddings.

For every post by user \(u\) with timestamp \(s_{i}\;(i=1,\dots,N)\) and embedding \(\mathbf{e}_{u}(s_{i})\in\mathbb{R}^{5}\), we compute a \emph{normalised time}

\[
\tau_{i}= \frac{s_{i}-t_{0}}{t_{\mathrm{end}}-t_{0}}\in[0,1],
\]

where \(t_{0}\) is 00:00 UTC on 17 Apr 2023 and \(t_{\mathrm{end}}\) is 23:59 UTC on 27 Oct 2023.  
A linear interpolant \(\mathbf{e}_{u}(\tau)\) is then fitted through the ordered pairs \((\tau_{i},\mathbf{e}_{u}(s_{i}))\).

Finally, we evaluate this interpolant at the 194 equally spaced grid points

\[
\tau_{g}= \frac{g}{193}\qquad(g=0,\dots,193),
\]

which correspond to \emph{noon of each calendar day}.  
The resulting sequence

\[
\bigl\{\mathbf{e}_{u}(\tau_{0}),\mathbf{e}_{u}(\tau_{1}),\dots,\mathbf{e}_{u}(\tau_{193})\bigr\}
\]

is the daily 5-dimensional trajectory for user \(u\).  
If a user has no post before a particular grid point, we carry the earliest observed embedding backward; similarly, we carry the final embedding forward beyond the user’s last post.

A weekly version of each trajectory is obtained by coordinate-wise averaging over non-overlapping 7-day windows; the final five days (23 – 27 Oct) are omitted because they do not complete a full week.

\subsubsection{PERMANOVA of Trajectory Pairs}

To answer RQ1 (whether the Increasing Toxicity Group follows a different topical path from its reference group) and RQ2 (likewise for the Decreasing Toxicity Group), we ran two distance-based permutational MANOVAs (PERMANOVA; \citeauthor{anderson_new_2001}\,\citeyear{anderson_new_2001}).  
Our test treats each user’s entire interpolated path as one multivariate observation. Formally, we set up the following hypothesis test for each trajectory pair:

\begin{enumerate}
    \item \textbf{H0:} The two group trajectories are \textit{not} significantly different in their sequence of positions in embedding space.

    \item \textbf{H1:} The two group trajectories are significantly different in their sequence of positions in embedding space.
\end{enumerate}

For user \(u\) we flatten the 5-dimensional trajectory obtained in the previous section into a single vector containing the embeddings from either the daily grid (\(T=194\)) or the weekly grid (\(T=27\)):

\[
\mathbf{x}_u =
\bigl[
\mathbf{e}_u(\tau_{0})^{\!\top},
\mathbf{e}_u(\tau_{1})^{\!\top},
\dots,
\mathbf{e}_u(\tau_{T-1})^{\!\top}
\bigr]^{\!\top}
\in\mathbb{R}^{5T}.
\]

Thus each comparison involves two sets of points in \(\mathbb{R}^{5T}\):  
Group A has \(n_A\) vectors \(\{\mathbf{x}_1,\dots,\mathbf{x}_{n_A}\}\), Group B has \(n_B\) vectors \(\{\mathbf{x}_{n_A+1},\dots,\mathbf{x}_{N}\}\) with \(N=n_A+n_B\). In our model \(n_A = n_B\) (718 vs.\ 718 for the Increasing Toxicity pair, 406 vs.\ 406 for the Decreasing Toxicity pair).

We use the ordinary Euclidean distance to form two sums of squares:
\[
\text{SS}_{\text{between}}
      = n_A\bigl\lVert\bar{\mathbf{x}}_A-\bar{\mathbf{x}}\bigr\rVert_2^{2}
        + n_B\bigl\lVert\bar{\mathbf{x}}_B-\bar{\mathbf{x}}\bigr\rVert_2^{2},
\]
\[
\text{SS}_{\text{within}}
      = \sum_{u\in A}\lVert\mathbf{x}_u-\bar{\mathbf{x}}_A\rVert_2^{2}
       +\sum_{u\in B}\lVert\mathbf{x}_u-\bar{\mathbf{x}}_B\rVert_2^{2},
\]

where the overall mean is  
\(\displaystyle \bar{\mathbf{x}}
   =\frac{n_A\bar{\mathbf{x}}_A+n_B\bar{\mathbf{x}}_B}{N}\), and the group-mean vectors are  
\(\bar{\mathbf{x}}_A=\tfrac{1}{n_A}\sum_{u\in A}\mathbf{x}_u\) and  
\(\bar{\mathbf{x}}_B=\tfrac{1}{n_B}\sum_{u\in B}\mathbf{x}_u\).  
\(\text{SS}_{\text{between}}\) measures the separation of the two group centroids, whereas \(\text{SS}_{\text{within}}\) measures the dispersion of trajectories around their own group mean.

The PERMANOVA statistic, \emph{pseudo-\(F\)} is

\[
F=\frac{\text{SS}_{\text{between}}}{\text{SS}_{\text{within}}/(N-2)},
\]

with \(1\) and \(N-2\) nominal degrees of freedom.

To obtain a \(p\)-value at \(\alpha=0.01\) we generated a null distribution of \(F\) by randomly reshuffling the user-level group labels 4,999 times, following Anderson \shortcite{anderson_new_2001}.

\subsubsection{Average Trajectory by Group}

Building on the daily and weekly user 5D trajectories obtained through linear interpolation, we computed each group’s average trajectory by taking the Euclidean mean of the group’s user vectors at each time step. Formally, for a given group $G$ and day $d$, the group-average embedding is:
\[
\bar{\mathbf{e}}_G(d) = \frac{1}{|G|} \sum_{u \in G} \mathbf{e}_u(d),
\]
the coordinate-wise centroid of all users in $G$ on day $d$. In the similar fashion, for a given group $G$ and week $w$, the group-average embedding is:
\[
\bar{\mathbf{e}}_G(w) = \frac{1}{|G|} \sum_{u \in G} \mathbf{e}_u(w),
\]
the coordinate-wise centroid of all users in $G$ on week $w$. Because the embeddings have been projected into a 5-dimensional UMAP space, which preserves important distance relationships, we judged that using Euclidean averages to compute each group's average trajectory is conceptually valid. 

\subsubsection{Topic Classification of Average Trajectories}
To convert each group’s average trajectory as a sequence of interpretable topics, we applied a nearest-neighbor classifier to every timestep of the trajectory. In particular, we trained a 15-nearest-neighbor (KNN) classifier in the 5-dimensional UMAP embedding space, using cosine similarity as the distance metric for finding neighbors. The classifier’s training data consisted solely of posts that were assigned to one of the 157 discovered topics from our Topic Discovery step. 

This KNN model proved highly effective, achieving near-perfect performance on held-out test data (macro F1 = 99.93 and micro F1 = 99.95). We used this classifier to label each point along the average trajectories. As a result, each group’s average daily and weekly trajectory in embedding space was transformed into a sequence of interpretable topics.

\section{Results}

\subsection{Topic Discovery}

Our recursive HDBSCAN procedure uncovered 157 Level-2 subtopics that span the policy, humanitarian, economic, cultural, and conspiratorial aspects of the U.S. immigration conversation (Appendix C). 

\begin{itemize}
    \item \emph{Policy, law and enforcement.}  
          Classical policy arguments appear in topics such as \textit{``Supreme Court Immigration Rulings''} (Topic 29), \textit{``DACA''} (Topic 8), or \textit{``E-Verify''} (Topic 112). Toxicity in this bloc is generally moderate (40–50), indicating relatively civil, though partisan, legal argumentation. 
          
    \item \emph{Economic and labor frames.}  
          Topics that portray migrants as either vital labor (\textit{``Immigrant Work Ethic''}, Topic 87; \textit{``Labor Shortage and Immigration''}, Topic 115) or unwelcome competitors (\textit{``Hiring of Illegal Immigrants''}, Topic 117) show different levels of toxicity (mean scores 45 vs.\ 57), illustrating how the same economic lens can lead to opposing narratives.
          
    \item \emph{Humanitarian Support.}  
          Humanitarian narratives cluster at the low-toxicity extreme, such as \textit{``Support for Migrant Communities''} (Topic 59, toxicity 30.5) and \textit{``Immigration Support and Resources''} (Topic 78, toxicity 20.3), frequently co-occurring with resource coordination such as housing, legal aid or language access. 
          
    \item \emph{Hostile nationalism.}  
          The highest toxicity averages are concentrated in “threat” frames that depict immigration as criminal invasion or demographic subversion: \textit{``Replacement Migration Conspiracy''} (Topic 89), \textit{``Anti-Immigration White Nationalism''} (Topic 132), and \textit{``Trump’s Anti-Immigrant Rhetoric''} (Topic 94). Keywords here such as \textit{``invasion''} and \textit{``poisoning the blood''} reveal that these clusters are shaped by fear-based narratives and strong anti-immigrant hostility.
          
    \item \emph{Identity-focused debates.}  
          Several Level-2 clusters reveal narratives centered on identity-based group tensions, such as \textit{``Black Americans on Immigration''} (Topic 20) and \textit{``Black American vs.\ Immigrant Dynamics''} (Topic 135). These topics suggest that conflicts within minority communities also play a role in shaping discourse around the U.S. immigration issue.
\end{itemize}

Label collisions, such as the presence of three distinct \textit{``Illegal Immigration Debate''} clusters (Topics 66, 146, and 155), might appear redundant. However, close reading reveals subtle differences in stance. (Full descriptions of all topics are accessible at https://topic-immigration.onrender.com). Topic 66 focuses on disputes concerning the distinction between asylum seekers and illegal immigrants. Topic 146 centers on debates over the usage of the term ``illegal immigrants'', whereas Topic 155 primarily includes emotionally charged condemnations. These nuanced differences among seemingly similar topics demonstrate that they can be further divided by their rhetorical emphasis.

In four high-traffic Level 2 subtopics, our iterative clustering process revealed a third layer of subtopics. One illustrative example is \textit{``NYC Migrant Housing Crisis''} (Topic 92), which was split into three subtopics. Although these separate clusters were retained due to their coherence scores being statistically higher than that of their parent topic, their descriptions show substantial overlap. Therefore, the semantic distinctions between these Level 3 clusters may have been difficult to detect using our labeling approach.

\subsection{Topic Trajectory Analysis Results}

\subsubsection{PERMANOVA Test Results}

To answer RQ1 and RQ2, we used PERMANOVA to compare the average topic trajectories of the Increasing Toxicity Group and the Decreasing Toxicity Group to those of their respective toxicity-matched reference groups. 

\begin{table}[t]
\centering
\begin{tabular}{
    >{\centering\arraybackslash}p{1.2cm}   % Resolution
    >{\centering\arraybackslash}p{2.1cm}   % Group Pair
    >{\centering\arraybackslash}p{1.6cm}   % Pseudo-F
    >{\centering\arraybackslash}p{0.6cm}   % p
    >{\centering\arraybackslash}p{0.6cm}   % eta^2
}
\toprule
\textbf{Frequency} & \textbf{Group Pair} & \textbf{Pseudo-$\mathbf{F}$} & \textbf{$p$} & \textbf{$\boldsymbol{\eta}^{2}$} \\
\midrule
\multirow{2}{1.2cm}{\centering Daily}  & \makecell[c]{Increasing vs. \\ Reference}  & 1.898 & .023 & .001 \\ \cline{2-5}
        & \makecell[c]{Decreasing vs. \\ Reference}  & 1.424 & .045 & .002 \\[4pt] \hline
\multirow{2}{1.2cm}{\centering Weekly}  & \makecell[c]{Increasing vs.\\Reference}  & 2.211 & .023 & .002 \\ \cline{2-5}
        & \makecell[c]{Decreasing vs. \\Reference}  & 1.619 & .040 & .002 \\
\bottomrule
\end{tabular}
\caption{PERMANOVA results comparing 5-D topic trajectories of toxicity-changing groups with toxicity-matched reference groups at daily ($T{=}194$) and weekly ($T{=}27$) frequencies. $p$-values are based on 4,999 permutations.}
\label{tab:permanova}
\end{table}

Table \ref{tab:permanova} reports the results for both temporal granularities. All four comparisons reject the null hypothesis at the \(\alpha = 0.05\) level, answering our two research questions.

\begin{itemize}
    \item \textbf{RQ1 (Increasing Toxicity Group):} The Increasing Toxicity Group exhibited significantly different trajectories from its toxicity-matched reference group.
    
    \item \textbf{RQ2 (Decreasing Toxicity Users):} The Decreasing Toxicity Group also followed significantly different trajectories from its matched reference group.
\end{itemize}

The group effect sizes turned out to be small (\(\eta^{2}\le .002\)), indicating that group differences account for only 0.2\% of the total dispersion, which is unsurprising given the presence of other sources of variation driving topic trajectories. Also, consistent pattern across temporal frequencies underscores that the reliable trajectory separation is not due to high-frequency noise.

\subsubsection{Comparison of Trajectory Pairs}

In line with the statistical evidence (Table~\ref{tab:permanova}), we find clear qualitative differences in the average topic trajectories of user groups whose toxicity is increasing or decreasing compared to their respective stable counterparts. Below, we compare the weekly topic trajectories of the Increasing Toxicity Group versus its Reference Group (Table \ref{tab:qual_inc} in Appendix D), and of the Decreasing Toxicity Group versus its Reference Group (Table \ref{tab:qual_dec} in Appendix D), focusing on key transitions and divergences.

\paragraph{\textit{Increasing Toxicity Group vs. Reference Group}}
By comparing weekly topic trajectories shown in Table \ref{tab:qual_inc} (Appendix D), we observe the Increasing Toxicity Group diverging notably from its Reference Group around late summer. In August, the Increasing Toxicity Group’s average toxicity rises sharply (Figure \ref{group_tox}), coinciding with shifts to more threat-oriented topics. For example, in the week of 7/31–8/6, this group gravitated to \textit{``Secret Flights of Illegal Immigrants''} (Topic 39, toxicity 56.67), a conspiracy-tinged topic implying covert government actions, whereas the Reference Group remained on a more conventional \textit{``155. Illegal Immigration Debate''} (Topic 155, toxicity 56.78). A few weeks later, from August 28 to September 3, the Increasing Toxicity Group’s average trajectory shifted to \textit{``U.S. Immigration Concerns''} (Topic 147, toxicity 61.67) and stayed there for the next five weeks, reflecting heightened fears about immigration. 

By mid-September, Increasing Toxicity Group’s average toxicity surpassed the Reference Group’s (Figure \ref{group_tox}), aligning with their emphasis on threat narratives. In contrast, the Reference Group’s average trajectory during these weeks showed little departure from its April pattern, repeatedly returning to one of its most common topics, \textit{``Anti-Immigration Sentiment''} (Topic 127, toxicity 70.58). This pattern addresses RQ1, suggesting that users whose toxicity increased shifted from early, sometimes humanitarian topics such as \textit{``Temporary Protected Status for Immigrants''} (Topic 22, toxicity 30.18) to more alarmist themes, exemplified by \textit{``Immigration and National Security''} (Topic 138, toxicity 68.13), during the critical August–September period, whereas their Reference Group maintained a steadier trajectory.

\paragraph{\textit{Decreasing Toxicity Group vs. Reference Group}}
In contrast, the Decreasing Toxicity Group’s trajectory shows a shift toward less inflammatory topics (Table \ref{tab:qual_dec} in Appendix D). In late spring 2023, this group occasionally engaged with the highly toxic and emotive topic \textit{``Ilhan Omar Immigration Controversy''} (Topic 5, toxicity 73.93). However, by mid-summer the Decreasing Toxicity Group began to concentrate on procedure-focused discussions. During the week of July 3 to July 9 its average trajectory visited \textit{``Discrimination and Immigration Laws''} (Topic 37, toxicity 52.73), while the Reference Group showed a similar pattern with \textit{Merit-Based Immigration Debate''} (Topic 84, toxicity 43.18). Over the next three weeks the Decreasing Toxicity Group remained in comparatively low-toxicity areas, including \textit{``Legal vs Illegal Immigration''} (Topic 96, toxicity 47.28), \textit{``Asylum Seekers and Immigration''} (Topic 76, toxicity 47.88), and \textit{``Immigration Detention Criticism''} (Topic 72, toxicity 39.79).

Corresponding with these topical shifts, the Decreasing Toxicity Group’s average toxicity fell below that of the Reference Group by mid-July (Figure \ref{group_tox}) and generally remained lower thereafter. While the Reference Group’s average trajectory periodically visited the highly toxic topic \textit{``U.S. Immigration Politics and Controversy''} (Topic 23, toxicity 71.11), the Decreasing Toxicity Group lingered on policy- and legal-focused themes such as \textit{``Supreme Court Immigration Rulings''} (Topic 29, toxicity 31.79). This suggests that members of the Decreasing Toxicity Group increasingly framed immigration as an institutional issue to be resolved through formal processes rather than as a partisan battle. By contrast, the Reference Group continued to gravitate toward politically charged frames, exemplified by \textit{``Republican Rhetoric on Immigration''} (Topic 88, toxicity 52.67) and \textit{Immigration and Social Justice Issues''} (Topic 101, toxicity 69.61). This pattern addresses RQ2: users whose toxicity decreased followed a distinct topical trajectory, an overall pivot to policy-oriented discussions.

\section{Discussion}

\subsection {Contribution}

\subsubsection{Academic Contribution} First, this study bridges topic-centered and user-centered toxicity research. By integrating user-focused analyses with a robust subtopic discovery approach, we highlight how specific subtopics can be associated with distinct forms of toxic behavior. This addresses the gap between topic-oriented toxicity research and user-level behavioral studies, offering a more holistic view of the dynamics of incivility in social media discourse.

Second, we empirically examine the longitudinal co-evolution of topic engagement and toxic communication. By modeling users’ trajectories across subtopics over time, we demonstrate that rising or falling toxicity can be closely linked to shifts in their content focus. Instead of viewing users' toxicity as a fixed attribute, we show how subtopic trajectories can illuminate pathways leading to intensified hostility or redirecting discussions toward policy debates.

Lastly, this study advances methodology on two fronts. First, we pair instruction-based embeddings with recursive HDBSCAN and coherence-guided hierarchical merging to extract stable, interpretable subtopics from massive social-media corpora about complex social issues. Second, our trajectory-analysis framework — which treats each user’s posting history as a continuous path through embedding space and compares groups using PERMANOVA — offers a replicable template for future research on the dynamics of social media discourse.

\subsubsection{Implications on Policy Communication}
Toxic discourse on immigration can foster societal divisions and spread misinformation about immigrants’ economic and social contributions. However, it would be reductive to label large groups of users as ``toxic'' in a static sense, as this would obscure the more nuanced narrative shaping public opinion.
Negativity in a policy discussion often arises in response to specific triggers—such as unmet policy needs, perceived inequities, or anxieties about future scenarios. Some individuals may start with moderate concerns about an immigration policy but escalate toward more toxic language when they feel their worries are dismissed or when policy outcomes fail to materialize as promised. In that sense, negativity can be a dynamic signal of policy-related frustration or dissatisfaction. This study can help policy communicators recognize that rising toxicity may be intertwined with specific policy-related frustrations, enabling them to devise strategies to de-escalate hostility and steer conversations back toward solution-oriented dialogue.

\subsection{Limitations and Ethical Considerations}
The dataset used for this study consists of posts in English, excluding discourse from users who primarily communicate in other languages. This introduces a potential bias by restricting the analysis to English-speaking populations, which may not fully represent broader discourse on U.S. immigration.

Second, our workflow extensively leveraged large language models, including zero-shot filtering of relevant posts, automated toxicity scoring, and LLM-based topic coherence scoring and labeling. While these models significantly reduce the cost of processing millions of posts, they also introduce errors and may reflect linguistic or cultural biases. The future study pipeline should therefore incorporate more systematic audits of misclassifications and robustness checks using alternative models. Expanding the proportion of human-validated ground truth at each stage will also help address the issue of bias.

The potential negative societal impacts of this research include reinforcing the stigmatization of certain user groups. To mitigate this risk, all identifiers, including user IDs, remain within the scope of the study and are not publicly disclosed, ensuring anonymity. No additional personally identifiable information was collected or used in the research. Furthermore, the results are framed to emphasize understanding and addressing toxicity constructively, rather than singling out specific user groups.

Potential misuse of this work includes justifying censorship, discrimination, or punitive measures against specific user groups based on their association with certain topics or levels of toxicity. To address these concerns, we reiterate that the purpose of this research is to foster an understanding of discourse dynamics on social issues and their implications for policy communications.

Finally, although users and topical factors may effectively inform us insights in creating healthy conversation and make potentially proactive interventions, there exist also other factors such as platform-level factors (e.g., \cite{dicicco_toxicity_2020}) at play here. Future research should explore further the connection between toxicity patterns, its antecedents (e.g., \cite{shen_viral_2020}), and outcomes.

\subsection{Conclusion}
In summary, our large-scale, six-month analysis of U.S.-immigration discourse on X shows that toxicity is tightly intertwined with how users move through the issue space. When people’s toxicity climbs, their posts shift toward alarmist and conspiracy-tinged subtopics, whereas falling toxicity is paired with a pivot to procedure- and policy-oriented themes. These findings highlight the fluid, topic-dependent character of incivility around contentious social issues and point to the limits of treating ``toxic users'' as a fixed category. In addition, the analytical toolkit we introduce, which features hierarchical topic discovery and a topic trajectory analysis pipeline, provides a replicable template for future research and for practical efforts to cultivate healthier online public spheres.

\bibliography{anonymous-submission-latex-2025}

\section{Paper Checklist}

\begin{enumerate}

\item For most authors...
\begin{enumerate}
    \item  Would answering this research question advance science without violating social contracts, such as violating privacy norms, perpetuating unfair profiling, exacerbating the socio-economic divide, or implying disrespect to societies or cultures?
    \answerYes{Yes}
  \item Do your main claims in the abstract and introduction accurately reflect the paper's contributions and scope?
    \answerYes{Yes}
   \item Do you clarify how the proposed methodological approach is appropriate for the claims made? 
    \answerYes{Yes}
   \item Do you clarify what are possible artifacts in the data used, given population-specific distributions?
    \answerYes{Yes, see the Data and Limitations and Ethical Considerations section.}
  \item Did you describe the limitations of your work?
    \answerYes{Yes, see the Limitations and Ethical Considerations section.}
  \item Did you discuss any potential negative societal impacts of your work?
    \answerYes{Yes, see the Limitations and Ethical Considerations section.}
      \item Did you discuss any potential misuse of your work?
    \answerYes{Yes, see the Limitations and Ethical Considerations section.}
    \item Did you describe steps taken to prevent or mitigate potential negative outcomes of the research, such as data and model documentation, data anonymization, responsible release, access control, and the reproducibility of findings?
        \answerYes{Yes, see the Limitations and Ethical Considerations section.}
  \item Have you read the ethics review guidelines and ensured that your paper conforms to them?
    \answerYes{Yes}
\end{enumerate}

\item Additionally, if your study involves hypotheses testing...
\begin{enumerate}
  \item Did you clearly state the assumptions underlying all theoretical results?
    \answerNA{NA}
  \item Have you provided justifications for all theoretical results?
    \answerNA{NA}
  \item Did you discuss competing hypotheses or theories that might challenge or complement your theoretical results?
    \answerNA{NA}
  \item Have you considered alternative mechanisms or explanations that might account for the same outcomes observed in your study?
    \answerNA{NA}
  \item Did you address potential biases or limitations in your theoretical framework?
    \answerNA{NA}
  \item Have you related your theoretical results to the existing literature in social science?
    \answerNA{NA}
  \item Did you discuss the implications of your theoretical results for policy, practice, or further research in the social science domain?
    \answerNA{NA}
\end{enumerate}

\item Additionally, if you are including theoretical proofs...
\begin{enumerate}
  \item Did you state the full set of assumptions of all theoretical results?
    \answerNA{NA}
	\item Did you include complete proofs of all theoretical results?
    \answerNA{NA}
\end{enumerate}

\item Additionally, if you ran machine learning experiments...
\begin{enumerate}
  \item Did you include the code, data, and instructions needed to reproduce the main experimental results (either in the supplemental material or as a URL)?
    \answerYes{Yes, detailed instructions for the analysis are provided in the Methodologies section and Appendix. However, the social media data will not be released due to privacy concerns.}
  \item Did you specify all the training details (e.g., data splits, hyperparameters, how they were chosen)?
    \answerYes{Yes, see the Methodologies section.}
     \item Did you report error bars (e.g., with respect to the random seed after running experiments multiple times)?
    \answerYes{Yes, see the Methodologies section.}
	\item Did you include the total amount of compute and the type of resources used (e.g., type of GPUs, internal cluster, or cloud provider)?
    \answerYes{Yes, see the Methodologies section.}
     \item Do you justify how the proposed evaluation is sufficient and appropriate to the claims made? 
    \answerYes{Yes, see the Methodologies section.}
     \item Do you discuss what is ``the cost`` of misclassification and fault (in)tolerance?
    \answerYes{Yes, see the Methodologies section.}
  
\end{enumerate}

\item Additionally, if you are using existing assets (e.g., code, data, models) or curating/releasing new assets, \textbf{without compromising anonymity}...
\begin{enumerate}
  \item If your work uses existing assets, did you cite the creators?
    \answerYes{Yes, see the Data and Methodologies section.}
  \item Did you mention the license of the assets?
    \answerYes{Yes, see the Methodologies section.}
  \item Did you include any new assets in the supplemental material or as a URL?
    \answerNA{NA}
  \item Did you discuss whether and how consent was obtained from people whose data you're using/curating?
    \answerNo{No, because the dataset used in this study consists of publicly available posts from X. As such, no explicit consent was obtained from individual users.}
  \item Did you discuss whether the data you are using/curating contains personally identifiable information or offensive content?
    \answerYes{Yes, see the Data, Methodologies and Limitations and Ethical Considerations section.}
\item If you are curating or releasing new datasets, did you discuss how you intend to make your datasets FAIR (see \citet{fair})?
\answerNA{NA}
\item If you are curating or releasing new datasets, did you create a Datasheet for the Dataset (see \citet{gebru2021datasheets})? 
\answerNA{NA}
\end{enumerate}

\item Additionally, if you used crowdsourcing or conducted research with human subjects, \textbf{without compromising anonymity}...
\begin{enumerate}
  \item Did you include the full text of instructions given to participants and screenshots?
    \answerNA{NA}
  \item Did you describe any potential participant risks, with mentions of Institutional Review Board (IRB) approvals?
    \answerNA{NA}
  \item Did you include the estimated hourly wage paid to participants and the total amount spent on participant compensation?
    \answerNA{NA}
   \item Did you discuss how data is stored, shared, and deidentified?
   \answerNA{NA}
\end{enumerate}

\appendix

\section{Appendix A: Prompt for Topic Coherence Rating}

\begin{listing}[htbp!]
\caption{Prompt for Topic Coherence Rating}
\label{lst:coherence-rating}
\begin{lstlisting}
Task Description:
You are a computational social scientist evaluating the coherence of a specific topic ("Topic A") ...

Evaluation Process:
1. Examine two sets of tweets:
   - In-topic examples: 30 randomly selected tweets classified as belonging to Topic A.
     {in_topic_examples}
   - Out-topic examples: 30 randomly selected tweets classified as NOT belonging to Topic A.
     {out_topic_examples}

2. Rate the coherence of Topic A on a 5-point Likert scale:
   - 5: Highly coherent.
   - 4: Moderately coherent.
   - 3: Neutral.
   - 2: Somewhat incoherent.
   - 1: Highly incoherent.

Note:
Relevance to U.S. immigration alone does not imply coherence. Evaluate distinctness of subtopics.

Output Format:
Coherence: {{coherence rate}}
\end{lstlisting}
\end{listing}

\end{enumerate}

\section{Appendix B: Prompt for Toxicity Detection}

\begin{listing}[htbp!]
\caption{Prompt for Toxicity Score Annotation}
\label{lst:toxicity-rating}
\begin{lstlisting}
Task Description:
You will be given a "text" in your language.
Your task is to assign a holistic toxicity score on a 5-point Likert scale to the text. 
Toxicity can take different forms, such as identity attack, bias, microaggression, violence, self-harm, insult, or sexual content.

Note:
- This dataset may contain extremely toxic content.
- Evaluate the text based on how toxic it would be perceived by a native speaker of the language.
- The text may be truncated.

Input:
Text: "{text}"

Output Format:
Respond with a single integer toxicity score [1-5]. Do not include anything else.
\end{lstlisting}
\end{listing}

\section{Appendix C: Detailed Topic Discovery Results}

\begin{table*}[htbp!]
\centering
\begin{tabular}{
  >{\centering\arraybackslash}m{1.2cm}
  >{\centering\arraybackslash}m{1.5cm}
  >{\centering\arraybackslash}m{1.2cm}
  >{\centering\arraybackslash}m{6.9cm}
  >{\centering\arraybackslash}m{1.5cm}
  >{\centering\arraybackslash}m{2.0cm}
}
\toprule
\textbf{\makecell{Lv.1 \\ Cluster}} & \textbf{\makecell{Lv.2 \\ Cluster}} & \textbf{\makecell{Lv.3 \\ Cluster}} &
\textbf{Label} & \textbf{Toxicity} & \textbf{Count} \\ 
\midrule

\multirow{2}{*}{\makecell{Outliers}} & \multirow{2}{*}{\makecell{}} & \multirow{2}{*}{\makecell{}} & \makecell{}  & \makecell{} & \makecell{94,558} \\
\cmidrule(lr){4-6}
&  &  & \multicolumn{3}{l}{\makecell[{{l}}]{}} \\
\midrule
\multirow{2}{*}{\makecell{1}} & \multirow{2}{*}{\makecell{}} & \multirow{2}{*}{\makecell{}} & \makecell{Anti-Immigrant Sentiment}  & \makecell{} & \makecell{29,902} \\
\cmidrule(lr){4-6}
&  &  & \multicolumn{3}{l}{\makecell[{{l}}]{The topic highlights hostile, divisive attitudes toward undocumented immigrants, \\ advocating stricter border control and portraying them as threats.}} \\
\midrule
\multirow{2}{*}{\makecell{2}} & \multirow{2}{*}{\makecell{}} & \multirow{2}{*}{\makecell{}} & \makecell{Immigrant Experiences and Stories}  & \makecell{} & \makecell{23,481} \\
\cmidrule(lr){4-6}
&  &  & \multicolumn{3}{l}{\makecell[{{l}}]{This topic highlights tweets sharing immigrant experiences, challenges, \\ triumphs, cultural diversity, and promoting empathy and understanding of \\ immigration.}} \\
\midrule
\multirow{2}{*}{\makecell{3}} & \multirow{2}{*}{\makecell{}} & \multirow{2}{*}{\makecell{}} & \makecell{Undocumented Immigration Issues}  & \makecell{} & \makecell{10,538} \\
\cmidrule(lr){4-6}
&  &  & \multicolumn{3}{l}{\makecell[{{l}}]{The topic addresses undocumented immigration in the U.S., highlighting \\ challenges, economic and societal impacts, and differing perspectives on policy.}} \\
\midrule
\multirow{2}{*}{\makecell{4}} & \multirow{2}{*}{\makecell{}} & \multirow{2}{*}{\makecell{}} & \makecell{Migration and Immigration Issues}  & \makecell{} & \makecell{14,685} \\
\cmidrule(lr){4-6}
&  &  & \multicolumn{3}{l}{\makecell[{{l}}]{This topic explores migration, its societal impact, policies, border control, \\ economic growth, social integration, and diverse perspectives on immigration.}} \\
\midrule
\multirow{2}{*}{\makecell{5}} & \multirow{2}{*}{\makecell{}} & \multirow{2}{*}{\makecell{}} & \makecell{European Migrant Crisis}  & \makecell{} & \makecell{3,726} \\
\cmidrule(lr){4-6}
&  &  & \multicolumn{3}{l}{\makecell[{{l}}]{The European migrant crisis involves large-scale migration to Europe, impacting \\ nations and sparking political, social, and community challenges.}} \\
\midrule
\multirow{2}{*}{\makecell{6}} & \multirow{2}{*}{\makecell{}} & \multirow{2}{*}{\makecell{}} & \makecell{Israeli-Palestinian Conflict}  & \makecell{} & \makecell{4,405} \\
\cmidrule(lr){4-6}
&  &  & \multicolumn{3}{l}{\makecell[{{l}}]{The topic focuses on the Israeli-Palestinian conflict, highlighting illegal \\ settlements, Palestinian rights, and criticism of U.S. support for Israel.}} \\
\midrule
\multirow{2}{*}{\makecell{7}} & \multirow{2}{*}{\makecell{}} & \multirow{2}{*}{\makecell{}} & \makecell{Anti-Immigrant Sentiment and Hate}  & \makecell{} & \makecell{4,930} \\
\cmidrule(lr){4-6}
&  &  & \multicolumn{3}{l}{\makecell[{{l}}]{This topic highlights xenophobia, bigotry, and negative rhetoric toward \\ immigrants, often involving derogatory language, generalizations, and political \\ divisiveness.}} \\
\midrule
\multirow{2}{*}{\makecell{8}} & \multirow{2}{*}{\makecell{}} & \multirow{2}{*}{\makecell{}} & \makecell{Immigration Fraud and Scams}  & \makecell{} & \makecell{2,352} \\
\cmidrule(lr){4-6}
&  &  & \multicolumn{3}{l}{\makecell[{{l}}]{The topic addresses U.S. immigration fraud, including fake marriages, document \\ forgery, scams, and calls for stricter enforcement and verification.}} \\
\midrule
\multirow{2}{*}{\makecell{9}} & \multirow{2}{*}{\makecell{}} & \multirow{2}{*}{\makecell{}} & \makecell{Economic Migration Debate}  & \makecell{} & \makecell{6,165} \\
\cmidrule(lr){4-6}
&  &  & \multicolumn{3}{l}{\makecell[{{l}}]{The topic discusses the polarized debate on economic migrants, their impact, \\ immigration policies, and the distinction from refugees.}} \\
\midrule
\multirow{2}{*}{\makecell{10}} & \multirow{2}{*}{\makecell{}} & \multirow{2}{*}{\makecell{}} & \makecell{Deportation of Undocumented Immigrants}  & \makecell{} & \makecell{2,019} \\
\cmidrule(lr){4-6}
&  &  & \multicolumn{3}{l}{\makecell[{{l}}]{The topic focuses on calls for deporting undocumented immigrants, featuring \\ divisive, urgent, and critical opinions on stricter immigration enforcement.}} \\
\midrule
\multirow{2}{*}{\makecell{11}} & \multirow{2}{*}{\makecell{}} & \multirow{2}{*}{\makecell{}} & \makecell{U.S. Immigration Debate}  & \makecell{} & \makecell{4,454,514} \\
\cmidrule(lr){4-6}
&  &  & \multicolumn{3}{l}{\makecell[{{l}}]{The U.S. immigration debate covers border control, illegal immigration, migrant \\ rights, policy impacts, and calls for comprehensive reform.}} \\
\midrule
\multirow{2}{*}{\makecell{}} & \multirow{2}{*}{\makecell{11-1 \\ (Topic 1)}} & \multirow{2}{*}{\makecell{}} & \makecell{Human Trafficking and Immigration}  & \makecell{64.92} & \makecell{1,126} \\
\cmidrule(lr){4-6}
&  &  & \multicolumn{3}{l}{\makecell[{{l}}]{The topic highlights human trafficking, illegal immigration, migrant \\ exploitation, border risks, and criticism of U.S. government policies.}} \\
\bottomrule
\end{tabular}
\caption{Detailed Topic Discovery Results 1}
\label{table}
\end{table*}

\begin{table*}[ht]
\centering
\begin{tabular}{
  >{\centering\arraybackslash}m{1.2cm}
  >{\centering\arraybackslash}m{1.5cm}
  >{\centering\arraybackslash}m{1.2cm}
  >{\centering\arraybackslash}m{6.9cm}
  >{\centering\arraybackslash}m{1.5cm}
  >{\centering\arraybackslash}m{2.0cm}
}
\toprule
\textbf{\makecell{Lv.1 \\ Cluster}} & \textbf{\makecell{Lv.2 \\ Cluster}} & \textbf{\makecell{Lv.3 \\ Cluster}} &
\textbf{Label} & \textbf{Toxicity} & \textbf{Count} \\ 
\midrule

\multirow{2}{*}{\makecell{}} & \multirow{2}{*}{\makecell{11-2 \\ (Topic 2)}} & \multirow{2}{*}{\makecell{}} & \makecell{Anti-Immigration Sentiment in the US}  & \makecell{63.10} & \makecell{1,536} \\
\cmidrule(lr){4-6}
&  &  & \multicolumn{3}{l}{\makecell[{{l}}]{The topic covers tweets opposing U.S. immigration, linking it to economic \\ issues, favoring Trump’s policies, and citing a related book.}} \\
\midrule
\multirow{2}{*}{\makecell{}} & \multirow{2}{*}{\makecell{11-3 \\ (Topic 3)}} & \multirow{2}{*}{\makecell{}} & \makecell{Immigration and Public Health Concerns}  & \makecell{64.32} & \makecell{20,664} \\
\cmidrule(lr){4-6}
&  &  & \multicolumn{3}{l}{\makecell[{{l}}]{The topic explores debates on U.S. immigration policies' impact on public \\ health, focusing on disease spread, vaccinations, and political discourse.}} \\
\midrule
\multirow{2}{*}{\makecell{}} & \multirow{2}{*}{\makecell{11-4 \\ (Topic 4)}} & \multirow{2}{*}{\makecell{}} & \makecell{Saudi Border Guard Atrocities}  & \makecell{57.48} & \makecell{2,992} \\
\cmidrule(lr){4-6}
&  &  & \multicolumn{3}{l}{\makecell[{{l}}]{The topic highlights alleged killings of African migrants by Saudi border guards \\ and critiques the U.S. response to these abuses.}} \\
\midrule
\multirow{2}{*}{\makecell{}} & \multirow{2}{*}{\makecell{11-5 \\ (Topic 5)}} & \multirow{2}{*}{\makecell{}} & \makecell{Ilhan Omar Immigration Controversy}  & \makecell{73.93} & \makecell{3,282} \\
\cmidrule(lr){4-6}
&  &  & \multicolumn{3}{l}{\makecell[{{l}}]{The topic involves allegations of immigration fraud against Ilhan Omar, sparking \\ polarized debates on her citizenship and calls for deportation.}} \\
\midrule
\multirow{2}{*}{\makecell{}} & \multirow{2}{*}{\makecell{11-6 \\ (Topic 6)}} & \multirow{2}{*}{\makecell{}} & \makecell{Civics Test for Voting}  & \makecell{50.33} & \makecell{2,379} \\
\cmidrule(lr){4-6}
&  &  & \multicolumn{3}{l}{\makecell[{{l}}]{The topic explores requiring U.S. citizens to pass a civics test to vote, \\ sparking debates on informed voting and barriers.}} \\
\midrule
\multirow{2}{*}{\makecell{}} & \multirow{2}{*}{\makecell{11-7 \\ (Topic 7)}} & \multirow{2}{*}{\makecell{}} & \makecell{Abuse of Migrant Populations}  & \makecell{67.13} & \makecell{1,105} \\
\cmidrule(lr){4-6}
&  &  & \multicolumn{3}{l}{\makecell[{{l}}]{The topic highlights allegations of sexual misconduct by law enforcement against \\ vulnerable migrants, exposing power dynamics and institutional failures.}} \\
\midrule
\multirow{2}{*}{\makecell{}} & \multirow{2}{*}{\makecell{11-8 \\ (Topic 8)}} & \multirow{2}{*}{\makecell{}} & \makecell{DACA Immigration Policy Debate}  & \makecell{35.90} & \makecell{2,394} \\
\cmidrule(lr){4-6}
&  &  & \multicolumn{3}{l}{\makecell[{{l}}]{The topic discusses the DACA program, its legality, impact on "Dreamers," and \\ the political debates surrounding U.S. immigration policy.}} \\
\midrule
\multirow{2}{*}{\makecell{}} & \multirow{2}{*}{\makecell{11-9 \\ (Topic 9)}} & \multirow{2}{*}{\makecell{}} & \makecell{Biden Immigration Loan Policies}  & \makecell{55.43} & \makecell{1,708} \\
\cmidrule(lr){4-6}
&  &  & \multicolumn{3}{l}{\makecell[{{l}}]{The topic discusses Biden administration policies pressuring banks to lend to \\ illegal immigrants, sparking debate over financial access and risks.}} \\
\midrule
\multirow{2}{*}{\makecell{}} & \multirow{2}{*}{\makecell{11-10 \\ (Topic 10)}} & \multirow{2}{*}{\makecell{}} & \makecell{Illegal Immigration in India}  & \makecell{68.56} & \makecell{1,218} \\
\cmidrule(lr){4-6}
&  &  & \multicolumn{3}{l}{\makecell[{{l}}]{The topic discusses concerns over illegal immigration in India, focusing on \\ impacts, government handling, and calls for stricter border control.}} \\
\midrule
\multirow{2}{*}{\makecell{}} & \multirow{2}{*}{\makecell{11-11 \\ (Topic 11)}} & \multirow{2}{*}{\makecell{}} & \makecell{Texas Migrant Shelter Attack}  & \makecell{59.35} & \makecell{2,954} \\
\cmidrule(lr){4-6}
&  &  & \multicolumn{3}{l}{\makecell[{{l}}]{An SUV intentionally rammed into migrants in Brownsville, Texas, sparking \\ outrage, debates on immigration policies, and calls for accountability.}} \\
\midrule
\multirow{2}{*}{\makecell{}} & \multirow{2}{*}{\makecell{11-12 \\ (Topic 12)}} & \multirow{2}{*}{\makecell{}} & \makecell{Illegal Immigrants as Police Officers}  & \makecell{62.20} & \makecell{4,245} \\
\cmidrule(lr){4-6}
&  &  & \multicolumn{3}{l}{\makecell[{{l}}]{The debate over allowing illegal immigrants to become police officers in \\ Illinois sparks concerns about law, security, and political polarization.}} \\
\midrule
\multirow{2}{*}{\makecell{}} & \multirow{2}{*}{\makecell{11-13 \\ (Topic 13)}} & \multirow{2}{*}{\makecell{}} & \makecell{Vatican's Stance on Immigration}  & \makecell{57.28} & \makecell{1,404} \\
\cmidrule(lr){4-6}
&  &  & \multicolumn{3}{l}{\makecell[{{l}}]{The topic critiques Pope Francis' stance on U.S. immigration, questioning the \\ Vatican's actions and accusing him of hypocrisy.}} \\
\midrule
\multirow{2}{*}{\makecell{}} & \multirow{2}{*}{\makecell{11-14 \\ (Topic 14)}} & \multirow{2}{*}{\makecell{}} & \makecell{Birthright Citizenship Debate}  & \makecell{57.51} & \makecell{12,634} \\
\cmidrule(lr){4-6}
&  &  & \multicolumn{3}{l}{\makecell[{{l}}]{The debate on U.S. birthright citizenship focuses on granting automatic \\ citizenship to children of undocumented immigrants, referencing the 14th \\ Amendment.}} \\
\bottomrule
\end{tabular}
\caption{Detailed Topic Discovery Results 2}
\label{table}
\end{table*}

\begin{table*}[ht]
\centering
\begin{tabular}{
  >{\centering\arraybackslash}m{1.2cm}
  >{\centering\arraybackslash}m{1.5cm}
  >{\centering\arraybackslash}m{1.2cm}
  >{\centering\arraybackslash}m{6.9cm}
  >{\centering\arraybackslash}m{1.5cm}
  >{\centering\arraybackslash}m{2.0cm}
}
\toprule
\textbf{\makecell{Lv.1 \\ Cluster}} & \textbf{\makecell{Lv.2 \\ Cluster}} & \textbf{\makecell{Lv.3 \\ Cluster}} &
\textbf{Label} & \textbf{Toxicity} & \textbf{Count} \\ 
\midrule

\multirow{2}{*}{\makecell{}} & \multirow{2}{*}{\makecell{}} & \multirow{2}{*}{\makecell{11-14-1}} & \makecell{Chain Migration Debate}  & \makecell{} & \makecell{1,108} \\
\cmidrule(lr){4-6}
&  &  & \multicolumn{3}{l}{\makecell[{{l}}]{The topic discusses chain migration, its role in U.S. immigration policy, and \\ debates over political hypocrisy and double standards.}} \\
\midrule
\multirow{2}{*}{\makecell{}} & \multirow{2}{*}{\makecell{}} & \multirow{2}{*}{\makecell{11-14-2}} & \makecell{Birthright Citizenship Debate}  & \makecell{} & \makecell{6,965} \\
\cmidrule(lr){4-6}
&  &  & \multicolumn{3}{l}{\makecell[{{l}}]{The topic discusses the debate over birthright citizenship for children of \\ undocumented immigrants, focusing on the 14th Amendment and immigration policy.}} \\
\midrule
\multirow{2}{*}{\makecell{}} & \multirow{2}{*}{\makecell{11-15 \\ (Topic 15)}} & \multirow{2}{*}{\makecell{}} & \makecell{Ted Cruz Immigration Hypocrisy}  & \makecell{62.98} & \makecell{2,227} \\
\cmidrule(lr){4-6}
&  &  & \multicolumn{3}{l}{\makecell[{{l}}]{Criticism of Senator Ted Cruz for anti-immigrant rhetoric, highlighting \\ perceived hypocrisy given his Cuban immigrant father's background.}} \\
\midrule
\multirow{2}{*}{\makecell{}} & \multirow{2}{*}{\makecell{11-16 \\ (Topic 16)}} & \multirow{2}{*}{\makecell{}} & \makecell{Satirical Views on Immigration Benefits}  & \makecell{69.99} & \makecell{1,633} \\
\cmidrule(lr){4-6}
&  &  & \multicolumn{3}{l}{\makecell[{{l}}]{This topic critiques U.S. immigration policies through satirical tweets, \\ claiming illegal immigrants receive unfair benefits unavailable to citizens.}} \\
\midrule
\multirow{2}{*}{\makecell{}} & \multirow{2}{*}{\makecell{11-17 \\ (Topic 17)}} & \multirow{2}{*}{\makecell{}} & \makecell{Driver's Licenses for Immigrants}  & \makecell{32.65} & \makecell{1,552} \\
\cmidrule(lr){4-6}
&  &  & \multicolumn{3}{l}{\makecell[{{l}}]{The topic discusses granting driver's licenses to undocumented immigrants in the \\ US, highlighting benefits, state policies, and political debates.}} \\
\midrule
\multirow{2}{*}{\makecell{}} & \multirow{2}{*}{\makecell{11-18 \\ (Topic 18)}} & \multirow{2}{*}{\makecell{}} & \makecell{Immigration and Voter Fraud}  & \makecell{62.31} & \makecell{57,998} \\
\cmidrule(lr){4-6}
&  &  & \multicolumn{3}{l}{\makecell[{{l}}]{The topic discusses claims of illegal immigration enabling voter fraud, \\ influencing elections, and sparking polarized debates on immigration and \\ election integrity.}} \\
\midrule
\multirow{2}{*}{\makecell{}} & \multirow{2}{*}{\makecell{11-19 \\ (Topic 19)}} & \multirow{2}{*}{\makecell{}} & \makecell{Replacement of American Citizens}  & \makecell{71.91} & \makecell{2,044} \\
\cmidrule(lr){4-6}
&  &  & \multicolumn{3}{l}{\makecell[{{l}}]{The topic discusses a conspiracy theory claiming Democrats intentionally replace \\ American citizens with immigrants to shift demographics and voting power.}} \\
\midrule
\multirow{2}{*}{\makecell{}} & \multirow{2}{*}{\makecell{11-20 \\ (Topic 20)}} & \multirow{2}{*}{\makecell{}} & \makecell{Black Americans on Immigration}  & \makecell{71.89} & \makecell{18,230} \\
\cmidrule(lr){4-6}
&  &  & \multicolumn{3}{l}{\makecell[{{l}}]{Black Americans express concerns about illegal immigration's impact on their \\ communities, citing economic competition, political neglect, and social \\ inequality.}} \\
\midrule
\multirow{2}{*}{\makecell{}} & \multirow{2}{*}{\makecell{11-21 \\ (Topic 21)}} & \multirow{2}{*}{\makecell{}} & \makecell{U.S. Immigration Deportation Policies}  & \makecell{35.85} & \makecell{1,576} \\
\cmidrule(lr){4-6}
&  &  & \multicolumn{3}{l}{\makecell[{{l}}]{The topic covers U.S. deportation policies, focusing on resumed flights, Mexico \\ agreements, and their political and humanitarian impacts.}} \\
\midrule
\multirow{2}{*}{\makecell{}} & \multirow{2}{*}{\makecell{11-22 \\ (Topic 22)}} & \multirow{2}{*}{\makecell{}} & \makecell{Temporary Protected Status for Immigrants}  & \makecell{30.18} & \makecell{2,226} \\
\cmidrule(lr){4-6}
&  &  & \multicolumn{3}{l}{\makecell[{{l}}]{The U.S. granted Temporary Protected Status to Venezuelan migrants, enabling \\ legal work, deportation protection, and sparking diverse stakeholder reactions.}} \\
\midrule
\multirow{2}{*}{\makecell{}} & \multirow{2}{*}{\makecell{11-23 \\ (Topic 23)}} & \multirow{2}{*}{\makecell{}} & \makecell{U.S. Immigration Politics and Controversy}  & \makecell{71.11} & \makecell{4,186} \\
\cmidrule(lr){4-6}
&  &  & \multicolumn{3}{l}{\makecell[{{l}}]{The topic critiques Governor Ron DeSantis' immigration policies, linking them to \\ broader issues of discrimination, human rights, and social justice.}} \\
\midrule
\multirow{2}{*}{\makecell{}} & \multirow{2}{*}{\makecell{11-24 \\ (Topic 24)}} & \multirow{2}{*}{\makecell{}} & \makecell{U.S. Immigration Policy Debate}  & \makecell{40.33} & \makecell{13,857} \\
\cmidrule(lr){4-6}
&  &  & \multicolumn{3}{l}{\makecell[{{l}}]{The topic discusses U.S. immigration policies, focusing on Title 42's \\ expiration, migrant influx, border impacts, and polarized opinions.}} \\
\midrule
\multirow{2}{*}{\makecell{}} & \multirow{2}{*}{\makecell{11-25 \\ (Topic 25)}} & \multirow{2}{*}{\makecell{}} & \makecell{Climate Change and Migration}  & \makecell{45.72} & \makecell{9,126} \\
\cmidrule(lr){4-6}
&  &  & \multicolumn{3}{l}{\makecell[{{l}}]{The topic explores how climate change drives migration, emphasizing policy \\ action, resource strain, and the need for climate justice.}} \\

\bottomrule
\end{tabular}
\caption{Detailed Topic Discovery Results 3}
\label{table}
\end{table*}

\begin{table*}[ht]
\centering
\begin{tabular}{
  >{\centering\arraybackslash}m{1.2cm}
  >{\centering\arraybackslash}m{1.5cm}
  >{\centering\arraybackslash}m{1.2cm}
  >{\centering\arraybackslash}m{6.9cm}
  >{\centering\arraybackslash}m{1.5cm}
  >{\centering\arraybackslash}m{2.0cm}
}
\toprule
\textbf{\makecell{Lv.1 \\ Cluster}} & \textbf{\makecell{Lv.2 \\ Cluster}} & \textbf{\makecell{Lv.3 \\ Cluster}} &
\textbf{Label} & \textbf{Toxicity} & \textbf{Count} \\ 
\midrule

\multirow{2}{*}{\makecell{}} & \multirow{2}{*}{\makecell{11-26 \\ (Topic 26)}} & \multirow{2}{*}{\makecell{}} & \makecell{Court Rulings on Immigration Policy}  & \makecell{34.02} & \makecell{1,931} \\
\cmidrule(lr){4-6}
&  &  & \multicolumn{3}{l}{\makecell[{{l}}]{The topic covers court rulings impacting Biden's immigration policies, sparking \\ debate over checks on power versus immigration reform and security.}} \\
\midrule
\multirow{2}{*}{\makecell{}} & \multirow{2}{*}{\makecell{11-27 \\ (Topic 27)}} & \multirow{2}{*}{\makecell{}} & \makecell{Support for Undocumented Students}  & \makecell{27.34} & \makecell{1,013} \\
\cmidrule(lr){4-6}
&  &  & \multicolumn{3}{l}{\makecell[{{l}}]{The topic focuses on support, resources, and advocacy for undocumented students \\ in U.S. higher education, promoting inclusivity and empowerment.}} \\
\midrule
\multirow{2}{*}{\makecell{}} & \multirow{2}{*}{\makecell{11-28 \\ (Topic 28)}} & \multirow{2}{*}{\makecell{}} & \makecell{U.S. Immigration Controversies}  & \makecell{61.77} & \makecell{6,882} \\
\cmidrule(lr){4-6}
&  &  & \multicolumn{3}{l}{\makecell[{{l}}]{The topic highlights controversies over U.S. immigration, focusing on family \\ separations, detention centers, policy criticism, and calls for accountability.}} \\
\midrule
\multirow{2}{*}{\makecell{}} & \multirow{2}{*}{\makecell{11-29 \\ (Topic 29)}} & \multirow{2}{*}{\makecell{}} & \makecell{Supreme Court Immigration Rulings}  & \makecell{31.79} & \makecell{1,686} \\
\cmidrule(lr){4-6}
&  &  & \multicolumn{3}{l}{\makecell[{{l}}]{The topic covers Supreme Court rulings on Biden administration immigration \\ policies, prioritizing deportations of public safety risks and border cases.}} \\
\midrule
\multirow{2}{*}{\makecell{}} & \multirow{2}{*}{\makecell{11-30 \\ (Topic 30)}} & \multirow{2}{*}{\makecell{}} & \makecell{Migrant Work Authorization}  & \makecell{34.93} & \makecell{2,561} \\
\cmidrule(lr){4-6}
&  &  & \multicolumn{3}{l}{\makecell[{{l}}]{The topic highlights the urgent need for expedited work permits for migrants, \\ emphasizing economic benefits and political advocacy for reforms.}} \\
\midrule
\multirow{2}{*}{\makecell{}} & \multirow{2}{*}{\makecell{11-31 \\ (Topic 31)}} & \multirow{2}{*}{\makecell{}} & \makecell{Healthcare for Undocumented Immigrants}  & \makecell{32.39} & \makecell{3,108} \\
\cmidrule(lr){4-6}
&  &  & \multicolumn{3}{l}{\makecell[{{l}}]{The topic discusses healthcare access for undocumented immigrants in the U.S., \\ including policies, challenges, and advocacy for universal healthcare.}} \\
\midrule
\multirow{2}{*}{\makecell{}} & \multirow{2}{*}{\makecell{11-32 \\ (Topic 32)}} & \multirow{2}{*}{\makecell{}} & \makecell{Immigration Moratorium Advocacy}  & \makecell{50.25} & \makecell{1,299} \\
\cmidrule(lr){4-6}
&  &  & \multicolumn{3}{l}{\makecell[{{l}}]{The topic discusses calls for an immigration moratorium in the U.S., citing job \\ competition, cultural shifts, and national security concerns.}} \\
\midrule
\multirow{2}{*}{\makecell{}} & \multirow{2}{*}{\makecell{11-33 \\ (Topic 33)}} & \multirow{2}{*}{\makecell{}} & \makecell{Reagan's Immigration Amnesty Policy}  & \makecell{50.24} & \makecell{1,235} \\
\cmidrule(lr){4-6}
&  &  & \multicolumn{3}{l}{\makecell[{{l}}]{The topic discusses Reagan's 1986 Immigration Reform, its amnesty for 3 million \\ immigrants, and its impact on current immigration debates.}} \\
\midrule
\multirow{2}{*}{\makecell{}} & \multirow{2}{*}{\makecell{11-34 \\ (Topic 34)}} & \multirow{2}{*}{\makecell{}} & \makecell{Amnesty for Illegal Immigrants}  & \makecell{52.52} & \makecell{2,115} \\
\cmidrule(lr){4-6}
&  &  & \multicolumn{3}{l}{\makecell[{{l}}]{The topic discusses the contentious debate on granting amnesty to illegal \\ immigrants, highlighting political divisions and varying public opinions.}} \\
\midrule
\multirow{2}{*}{\makecell{}} & \multirow{2}{*}{\makecell{11-35 \\ (Topic 35)}} & \multirow{2}{*}{\makecell{}} & \makecell{U.S. Immigration Policy Debate}  & \makecell{53.13} & \makecell{3,185} \\
\cmidrule(lr){4-6}
&  &  & \multicolumn{3}{l}{\makecell[{{l}}]{The topic discusses U.S. immigration policies, focusing on illegal immigration, \\ border control, economic impacts, and polarized opinions on leadership.}} \\
\midrule
\multirow{2}{*}{\makecell{}} & \multirow{2}{*}{\makecell{11-36 \\ (Topic 36)}} & \multirow{2}{*}{\makecell{}} & \makecell{Anti-Immigration Border Control}  & \makecell{63.36} & \makecell{3,036} \\
\cmidrule(lr){4-6}
&  &  & \multicolumn{3}{l}{\makecell[{{l}}]{The topic highlights opposition to illegal immigration, advocating border wall \\ construction, strict deportation, and aggressive border control for national \\ security.}} \\
\midrule
\multirow{2}{*}{\makecell{}} & \multirow{2}{*}{\makecell{11-37 \\ (Topic 37)}} & \multirow{2}{*}{\makecell{}} & \makecell{Discrimination and Immigration Laws}  & \makecell{52.73} & \makecell{1,948} \\
\cmidrule(lr){4-6}
&  &  & \multicolumn{3}{l}{\makecell[{{l}}]{This topic discusses polarized views on discrimination in U.S. immigration, \\ focusing on legality, race, ethnicity, and national origin.}} \\
\midrule
\multirow{2}{*}{\makecell{}} & \multirow{2}{*}{\makecell{11-38 \\ (Topic 38)}} & \multirow{2}{*}{\makecell{}} & \makecell{U.S. Immigration Controversy}  & \makecell{66.62} & \makecell{1,829} \\
\cmidrule(lr){4-6}
&  &  & \multicolumn{3}{l}{\makecell[{{l}}]{The topic addresses criticism of Texas Governor Greg Abbott's comment labeling \\ mass shooting victims as ``illegal immigrants,'' sparking immigration debate.}} \\

\bottomrule
\end{tabular}
\caption{Detailed Topic Discovery Results 4}
\label{table}
\end{table*}

\begin{table*}[ht]
\centering
\begin{tabular}{
  >{\centering\arraybackslash}m{1.2cm}
  >{\centering\arraybackslash}m{1.5cm}
  >{\centering\arraybackslash}m{1.2cm}
  >{\centering\arraybackslash}m{6.9cm}
  >{\centering\arraybackslash}m{1.5cm}
  >{\centering\arraybackslash}m{2.0cm}
}
\toprule
\textbf{\makecell{Lv.1 \\ Cluster}} & \textbf{\makecell{Lv.2 \\ Cluster}} & \textbf{\makecell{Lv.3 \\ Cluster}} &
\textbf{Label} & \textbf{Toxicity} & \textbf{Count} \\ 
\midrule

\multirow{2}{*}{\makecell{}} & \multirow{2}{*}{\makecell{11-39 \\ (Topic 39)}} & \multirow{2}{*}{\makecell{}} & \makecell{Secret Flights of Illegal Immigrants}  & \makecell{56.67} & \makecell{1,469} \\
\cmidrule(lr){4-6}
&  &  & \multicolumn{3}{l}{\makecell[{{l}}]{The topic discusses alleged secret flights of illegal immigrants under the Biden \\ administration, sparking concerns over transparency and national security.}} \\
\midrule
\multirow{2}{*}{\makecell{}} & \multirow{2}{*}{\makecell{11-40 \\ (Topic 40)}} & \multirow{2}{*}{\makecell{}} & \makecell{Christian Views on Immigration}  & \makecell{55.72} & \makecell{7,410} \\
\cmidrule(lr){4-6}
&  &  & \multicolumn{3}{l}{\makecell[{{l}}]{The topic explores Christianity's role in U.S. immigration debates on social \\ media, citing biblical arguments and highlighting political and moral divides.}} \\
\midrule
\multirow{2}{*}{\makecell{}} & \multirow{2}{*}{\makecell{11-41 \\ (Topic 41)}} & \multirow{2}{*}{\makecell{}} & \makecell{DeSantis Immigration Policy Debate}  & \makecell{42.36} & \makecell{1,263} \\
\cmidrule(lr){4-6}
&  &  & \multicolumn{3}{l}{\makecell[{{l}}]{The topic compares immigration policies of Ron DeSantis and Donald Trump, \\ highlighting debates on effectiveness, political implications, and Republican \\ stances.}} \\
\midrule
\multirow{2}{*}{\makecell{}} & \multirow{2}{*}{\makecell{11-42 \\ (Topic 42)}} & \multirow{2}{*}{\makecell{}} & \makecell{Migrant Shipwrecks and Public Response}  & \makecell{53.12} & \makecell{20,036} \\
\cmidrule(lr){4-6}
&  &  & \multicolumn{3}{l}{\makecell[{{l}}]{The topic highlights outrage over unequal media coverage and empathy for migrant \\ shipwrecks compared to wealthy individuals' tragedies.}} \\
\midrule
\multirow{2}{*}{\makecell{}} & \multirow{2}{*}{\makecell{11-43 \\ (Topic 43)}} & \multirow{2}{*}{\makecell{}} & \makecell{Immigration and Military Service}  & \makecell{60.51} & \makecell{4,762} \\
\cmidrule(lr){4-6}
&  &  & \multicolumn{3}{l}{\makecell[{{l}}]{The topic explores immigrants, including illegal ones, serving in the US \\ military to gain citizenship, sparking diverse political and social views.}} \\
\midrule
\multirow{2}{*}{\makecell{}} & \multirow{2}{*}{\makecell{11-44 \\ (Topic 44)}} & \multirow{2}{*}{\makecell{}} & \makecell{U.S. Immigration Policy History}  & \makecell{45.24} & \makecell{1,612} \\
\cmidrule(lr){4-6}
&  &  & \multicolumn{3}{l}{\makecell[{{l}}]{This topic discusses the historical impact, intentions, and debates surrounding \\ the U.S. Immigration and Nationality Act of 1965.}} \\
\midrule
\multirow{2}{*}{\makecell{}} & \multirow{2}{*}{\makecell{11-45 \\ (Topic 45)}} & \multirow{2}{*}{\makecell{}} & \makecell{Immigration Impact on Public Schools}  & \makecell{62.75} & \makecell{1,002} \\
\cmidrule(lr){4-6}
&  &  & \multicolumn{3}{l}{\makecell[{{l}}]{The topic highlights concerns about immigration's impact on U.S. public schools, \\ including strained resources, lowered standards, and financial burdens.}} \\
\midrule
\multirow{2}{*}{\makecell{}} & \multirow{2}{*}{\makecell{11-46 \\ (Topic 46)}} & \multirow{2}{*}{\makecell{}} & \makecell{Japan's Immigration Policies}  & \makecell{51.86} & \makecell{2,201} \\
\cmidrule(lr){4-6}
&  &  & \multicolumn{3}{l}{\makecell[{{l}}]{The topic discusses Japan's strict immigration policies, comparisons to Western \\ countries, and debates on cultural preservation versus economic needs.}} \\
\midrule
\multirow{2}{*}{\makecell{}} & \multirow{2}{*}{\makecell{11-47 \\ (Topic 47)}} & \multirow{2}{*}{\makecell{}} & \makecell{Illegal Economic Migration Debate}  & \makecell{60.51} & \makecell{1,358} \\
\cmidrule(lr){4-6}
&  &  & \multicolumn{3}{l}{\makecell[{{l}}]{The topic focuses on opposition to illegal economic migrants, highlighting \\ concerns about economy, crime, social benefits, and stricter border controls.}} \\
\midrule
\multirow{2}{*}{\makecell{}} & \multirow{2}{*}{\makecell{11-48 \\ (Topic 48)}} & \multirow{2}{*}{\makecell{}} & \makecell{Chinese Immigration and National Security}  & \makecell{63.77} & \makecell{2,773} \\
\cmidrule(lr){4-6}
&  &  & \multicolumn{3}{l}{\makecell[{{l}}]{Concerns about Chinese nationals crossing the U.S.-Mexico border, with \\ speculation about espionage, security threats, and calls for government action.}} \\
\midrule
\multirow{2}{*}{\makecell{}} & \multirow{2}{*}{\makecell{11-49 \\ (Topic 49)}} & \multirow{2}{*}{\makecell{}} & \makecell{Migrant Students in US Schools}  & \makecell{35.34} & \makecell{1,044} \\
\cmidrule(lr){4-6}
&  &  & \multicolumn{3}{l}{\makecell[{{l}}]{The topic highlights challenges of migrant students in US schools, focusing on \\ resources, emotional struggles, and local government responses.}} \\
\midrule
\multirow{2}{*}{\makecell{}} & \multirow{2}{*}{\makecell{11-50 \\ (Topic 50)}} & \multirow{2}{*}{\makecell{}} & \makecell{U.S. Border Wall Debate}  & \makecell{54.32} & \makecell{26,717} \\
\cmidrule(lr){4-6}
&  &  & \multicolumn{3}{l}{\makecell[{{l}}]{The topic discusses the polarized debate over the U.S.-Mexico border wall, \\ focusing on its effectiveness, cost, and political implications.}} \\
\midrule
\multirow{2}{*}{\makecell{}} & \multirow{2}{*}{\makecell{11-51 \\ (Topic 51)}} & \multirow{2}{*}{\makecell{}} & \makecell{Migrant Child Death Controversy}  & \makecell{55.62} & \makecell{1,472} \\
\cmidrule(lr){4-6}
&  &  & \multicolumn{3}{l}{\makecell[{{l}}]{The death of a 3-year-old migrant during Texas' busing program sparks outrage, \\ criticism of Governor Abbott, and immigration policy debates.}} \\

\bottomrule
\end{tabular}
\caption{Detailed Topic Discovery Results 5}
\label{table}
\end{table*}

\begin{table*}[ht]
\centering
\begin{tabular}{
  >{\centering\arraybackslash}m{1.2cm}
  >{\centering\arraybackslash}m{1.5cm}
  >{\centering\arraybackslash}m{1.2cm}
  >{\centering\arraybackslash}m{6.9cm}
  >{\centering\arraybackslash}m{1.5cm}
  >{\centering\arraybackslash}m{2.0cm}
}
\toprule
\textbf{\makecell{Lv.1 \\ Cluster}} & \textbf{\makecell{Lv.2 \\ Cluster}} & \textbf{\makecell{Lv.3 \\ Cluster}} &
\textbf{Label} & \textbf{Toxicity} & \textbf{Count} \\ 
\midrule

\multirow{2}{*}{\makecell{}} & \multirow{2}{*}{\makecell{11-52 \\ (Topic 52)}} & \multirow{2}{*}{\makecell{}} & \makecell{Free College for Illegal Immigrants}  & \makecell{56.25} & \makecell{1,508} \\
\cmidrule(lr){4-6}
&  &  & \multicolumn{3}{l}{\makecell[{{l}}]{The topic discusses controversy over U.S. states offering free college tuition \\ and benefits to undocumented immigrants, sparking fairness debates.}} \\
\midrule
\multirow{2}{*}{\makecell{}} & \multirow{2}{*}{\makecell{11-53 \\ (Topic 53)}} & \multirow{2}{*}{\makecell{}} & \makecell{UK Immigration to Rwanda Policy}  & \makecell{57.96} & \makecell{1,109} \\
\cmidrule(lr){4-6}
&  &  & \multicolumn{3}{l}{\makecell[{{l}}]{The UK's plan to send migrants to Rwanda sparks debate on immigration, human \\ rights, and its broader migration policies.}} \\
\midrule
\multirow{2}{*}{\makecell{}} & \multirow{2}{*}{\makecell{11-54 \\ (Topic 54)}} & \multirow{2}{*}{\makecell{}} & \makecell{Anti-Immigration Sentiment}  & \makecell{66.81} & \makecell{94,522} \\
\cmidrule(lr){4-6}
&  &  & \multicolumn{3}{l}{\makecell[{{l}}]{The topic highlights opposition to illegal immigration in the U.S., featuring \\ negative, xenophobic rhetoric, policy criticism, and misinformation.}} \\
\midrule
\multirow{2}{*}{\makecell{}} & \multirow{2}{*}{\makecell{11-55 \\ (Topic 55)}} & \multirow{2}{*}{\makecell{}} & \makecell{Refugee vs Illegal Immigrant Debate}  & \makecell{53.09} & \makecell{1,923} \\
\cmidrule(lr){4-6}
&  &  & \multicolumn{3}{l}{\makecell[{{l}}]{The topic highlights the distinction between refugees and illegal immigrants, \\ emphasizing legal, humanitarian differences, and politically charged debates.}} \\
\midrule
\multirow{2}{*}{\makecell{}} & \multirow{2}{*}{\makecell{11-56 \\ (Topic 56)}} & \multirow{2}{*}{\makecell{}} & \makecell{Undocumented Immigrants and Taxes}  & \makecell{51.28} & \makecell{3,505} \\
\cmidrule(lr){4-6}
&  &  & \multicolumn{3}{l}{\makecell[{{l}}]{The topic discusses undocumented immigrants' significant tax contributions, \\ countering misconceptions and highlighting their economic impact in the U.S.}} \\
\midrule
\multirow{2}{*}{\makecell{}} & \multirow{2}{*}{\makecell{11-57 \\ (Topic 57)}} & \multirow{2}{*}{\makecell{}} & \makecell{Migrant Child Labor Exploitation}  & \makecell{57.58} & \makecell{3,242} \\
\cmidrule(lr){4-6}
&  &  & \multicolumn{3}{l}{\makecell[{{l}}]{The topic highlights migrant child labor exploitation in hazardous jobs, \\ inadequate protections, and calls for stricter labor laws and standards.}} \\
\midrule
\multirow{2}{*}{\makecell{}} & \multirow{2}{*}{\makecell{11-58 \\ (Topic 58)}} & \multirow{2}{*}{\makecell{}} & \makecell{U.S. Immigration Court Backlog}  & \makecell{38.20} & \makecell{1,562} \\
\cmidrule(lr){4-6}
&  &  & \multicolumn{3}{l}{\makecell[{{l}}]{The U.S. immigration court system faces significant case backlogs, prompting \\ calls for more judges, funding, and streamlined processing for fairness.}} \\
\midrule
\multirow{2}{*}{\makecell{}} & \multirow{2}{*}{\makecell{11-59 \\ (Topic 59)}} & \multirow{2}{*}{\makecell{}} & \makecell{Support for Migrant Communities}  & \makecell{30.47} & \makecell{1,468} \\
\cmidrule(lr){4-6}
&  &  & \multicolumn{3}{l}{\makecell[{{l}}]{Tweets express support for migrants, highlighting aid efforts, essential \\ services, and advocacy for their rights and well-being.}} \\
\midrule
\multirow{2}{*}{\makecell{}} & \multirow{2}{*}{\makecell{11-60 \\ (Topic 60)}} & \multirow{2}{*}{\makecell{}} & \makecell{Green Card Backlog Issues}  & \makecell{50.05} & \makecell{2,462} \\
\cmidrule(lr){4-6}
&  &  & \multicolumn{3}{l}{\makecell[{{l}}]{Legal immigrants, especially Indians, face green card delays due to backlogs and \\ caps, sparking calls for U.S. immigration reforms.}} \\
\midrule
\multirow{2}{*}{\makecell{}} & \multirow{2}{*}{\makecell{11-61 \\ (Topic 61)}} & \multirow{2}{*}{\makecell{}} & \makecell{Human Smuggling and Border Control}  & \makecell{42.10} & \makecell{1,632} \\
\cmidrule(lr){4-6}
&  &  & \multicolumn{3}{l}{\makecell[{{l}}]{The topic focuses on U.S. immigration enforcement, highlighting human smuggling \\ incidents, law enforcement efforts, and border security challenges.}} \\
\midrule
\multirow{2}{*}{\makecell{}} & \multirow{2}{*}{\makecell{11-62 \\ (Topic 62)}} & \multirow{2}{*}{\makecell{}} & \makecell{Migrant Deaths and Border Crisis}  & \makecell{49.40} & \makecell{2,898} \\
\cmidrule(lr){4-6}
&  &  & \multicolumn{3}{l}{\makecell[{{l}}]{The topic highlights migrant deaths, inhumane conditions, border crisis, \\ government inaction, and the urgent need for immigration reform.}} \\
\midrule
\multirow{2}{*}{\makecell{}} & \multirow{2}{*}{\makecell{11-63 \\ (Topic 63)}} & \multirow{2}{*}{\makecell{}} & \makecell{Criticism of US Immigration Policies}  & \makecell{67.11} & \makecell{7,850} \\
\cmidrule(lr){4-6}
&  &  & \multicolumn{3}{l}{\makecell[{{l}}]{Criticism of Texas Governor Greg Abbott's immigration policies, including razor \\ wire use and migrant pushbacks, deemed inhumane and cruel.}} \\
\midrule
\multirow{2}{*}{\makecell{}} & \multirow{2}{*}{\makecell{11-64 \\ (Topic 64)}} & \multirow{2}{*}{\makecell{}} & \makecell{Expats vs Immigrants}  & \makecell{46.36} & \makecell{2,099} \\
\cmidrule(lr){4-6}
&  &  & \multicolumn{3}{l}{\makecell[{{l}}]{The topic examines the double standard in using "expats" for Western migrants \\ and "immigrants" for others, highlighting bias and inconsistency.}} \\

\bottomrule
\end{tabular}
\caption{Detailed Topic Discovery Results 6}
\label{table}
\end{table*}

\begin{table*}[ht]
\centering
\begin{tabular}{
  >{\centering\arraybackslash}m{1.2cm}
  >{\centering\arraybackslash}m{1.5cm}
  >{\centering\arraybackslash}m{1.2cm}
  >{\centering\arraybackslash}m{6.9cm}
  >{\centering\arraybackslash}m{1.5cm}
  >{\centering\arraybackslash}m{2.0cm}
}
\toprule
\textbf{\makecell{Lv.1 \\ Cluster}} & \textbf{\makecell{Lv.2 \\ Cluster}} & \textbf{\makecell{Lv.3 \\ Cluster}} &
\textbf{Label} & \textbf{Toxicity} & \textbf{Count} \\ 
\midrule

\multirow{2}{*}{\makecell{}} & \multirow{2}{*}{\makecell{11-65 \\ (Topic 65)}} & \multirow{2}{*}{\makecell{}} & \makecell{The Great Migration}  & \makecell{38.20} & \makecell{1,820} \\
\cmidrule(lr){4-6}
&  &  & \multicolumn{3}{l}{\makecell[{{l}}]{The Great Migration saw 6 million African Americans move from the South \\ (1915–1970), shaping history, arts, and social justice.}} \\
\midrule
\multirow{2}{*}{\makecell{}} & \multirow{2}{*}{\makecell{11-66 \\ (Topic 66)}} & \multirow{2}{*}{\makecell{}} & \makecell{Illegal Immigration Debate}  & \makecell{62.61} & \makecell{1,096} \\
\cmidrule(lr){4-6}
&  &  & \multicolumn{3}{l}{\makecell[{{l}}]{The topic discusses polarized views on U.S. illegal immigration, debating \\ distinctions between refugees and illegal immigrants and their impact.}} \\
\midrule
\multirow{2}{*}{\makecell{}} & \multirow{2}{*}{\makecell{11-67 \\ (Topic 67)}} & \multirow{2}{*}{\makecell{}} & \makecell{Enforcement of Immigration Laws}  & \makecell{49.11} & \makecell{6,612} \\
\cmidrule(lr){4-6}
&  &  & \multicolumn{3}{l}{\makecell[{{l}}]{The topic highlights frustration over perceived lack of immigration law \\ enforcement under Biden, with calls for stricter border control.}} \\
\midrule
\multirow{2}{*}{\makecell{}} & \multirow{2}{*}{\makecell{11-68 \\ (Topic 68)}} & \multirow{2}{*}{\makecell{}} & \makecell{US-Mexico Border Crisis}  & \makecell{39.26} & \makecell{1,252} \\
\cmidrule(lr){4-6}
&  &  & \multicolumn{3}{l}{\makecell[{{l}}]{The topic highlights the migrant surge at the US-Mexico border, straining \\ resources and prompting concerns over a local crisis.}} \\
\midrule
\multirow{2}{*}{\makecell{}} & \multirow{2}{*}{\makecell{11-69 \\ (Topic 69)}} & \multirow{2}{*}{\makecell{}} & \makecell{Immigration and Welfare Benefits}  & \makecell{56.20} & \makecell{6,321} \\
\cmidrule(lr){4-6}
&  &  & \multicolumn{3}{l}{\makecell[{{l}}]{The topic discusses U.S. immigration's impact on welfare, with concerns about \\ economic strain and debates on immigrants' contributions.}} \\
\midrule
\multirow{2}{*}{\makecell{}} & \multirow{2}{*}{\makecell{11-70 \\ (Topic 70)}} & \multirow{2}{*}{\makecell{}} & \makecell{DeSantis' Migrant Relocation Controversy}  & \makecell{52.74} & \makecell{18,930} \\
\cmidrule(lr){4-6}
&  &  & \multicolumn{3}{l}{\makecell[{{l}}]{The topic covers criticism of Florida Governor Ron DeSantis for relocating \\ migrants, facing accusations of exploitation, and calls for accountability.}} \\
\midrule
\multirow{2}{*}{\makecell{}} & \multirow{2}{*}{\makecell{11-71 \\ (Topic 71)}} & \multirow{2}{*}{\makecell{}} & \makecell{U.S.-Mexico Border Disputes}  & \makecell{33.66} & \makecell{1,493} \\
\cmidrule(lr){4-6}
&  &  & \multicolumn{3}{l}{\makecell[{{l}}]{The topic covers Texas' use of floating barriers in the Rio Grande, sparking \\ legal, humanitarian, and political disputes with the federal government.}} \\
\midrule
\multirow{2}{*}{\makecell{}} & \multirow{2}{*}{\makecell{11-72 \\ (Topic 72)}} & \multirow{2}{*}{\makecell{}} & \makecell{Immigration Detention Criticism}  & \makecell{39.79} & \makecell{1,917} \\
\cmidrule(lr){4-6}
&  &  & \multicolumn{3}{l}{\makecell[{{l}}]{The topic highlights criticism of the U.S. immigration detention system, \\ focusing on inhumane conditions, private prisons, and calls for reform.}} \\
\midrule
\multirow{2}{*}{\makecell{}} & \multirow{2}{*}{\makecell{11-73 \\ (Topic 73)}} & \multirow{2}{*}{\makecell{}} & \makecell{Refugee vs Economic Migrant Debate}  & \makecell{55.81} & \makecell{2,777} \\
\cmidrule(lr){4-6}
&  &  & \multicolumn{3}{l}{\makecell[{{l}}]{The topic debates distinguishing refugees from economic migrants, focusing on \\ skepticism of refugee claims and calls for stricter immigration controls.}} \\
\midrule
\multirow{2}{*}{\makecell{}} & \multirow{2}{*}{\makecell{11-74 \\ (Topic 74)}} & \multirow{2}{*}{\makecell{}} & \makecell{Economic Migrants vs Asylum Seekers}  & \makecell{56.06} & \makecell{1,860} \\
\cmidrule(lr){4-6}
&  &  & \multicolumn{3}{l}{\makecell[{{l}}]{The topic highlights skepticism about asylum claims, emphasizing the need to \\ differentiate genuine asylum seekers from economic migrants.}} \\
\midrule
\multirow{2}{*}{\makecell{}} & \multirow{2}{*}{\makecell{11-75 \\ (Topic 75)}} & \multirow{2}{*}{\makecell{}} & \makecell{Anti-Immigration Sentiment}  & \makecell{61.72} & \makecell{2,193} \\
\cmidrule(lr){4-6}
&  &  & \multicolumn{3}{l}{\makecell[{{l}}]{The topic highlights anti-immigration sentiment in the U.S., criticizing illegal \\ immigration, asylum seekers, and advocating stricter border control policies.}} \\
\midrule
\multirow{2}{*}{\makecell{}} & \multirow{2}{*}{\makecell{11-76 \\ (Topic 76)}} & \multirow{2}{*}{\makecell{}} & \makecell{Asylum Seekers and Immigration}  & \makecell{47.88} & \makecell{12,369} \\
\cmidrule(lr){4-6}
&  &  & \multicolumn{3}{l}{\makecell[{{l}}]{The topic highlights asylum seekers' legal rights, distinguishing them from \\ illegal immigrants, and addresses misconceptions through informative debates.}} \\
\midrule
\multirow{2}{*}{\makecell{}} & \multirow{2}{*}{\makecell{11-77 \\ (Topic 77)}} & \multirow{2}{*}{\makecell{}} & \makecell{Racist Immigration Rhetoric}  & \makecell{67.84} & \makecell{1,121} \\
\cmidrule(lr){4-6}
&  &  & \multicolumn{3}{l}{\makecell[{{l}}]{This topic examines racism in U.S. immigration debates, focusing on assumptions \\ about Latin American immigrants and the impact of racist rhetoric.}} \\

\bottomrule
\end{tabular}
\caption{Detailed Topic Discovery Results 7}
\label{table}
\end{table*}

\begin{table*}[ht]
\centering
\begin{tabular}{
  >{\centering\arraybackslash}m{1.2cm}
  >{\centering\arraybackslash}m{1.5cm}
  >{\centering\arraybackslash}m{1.2cm}
  >{\centering\arraybackslash}m{6.9cm}
  >{\centering\arraybackslash}m{1.5cm}
  >{\centering\arraybackslash}m{2.0cm}
}
\toprule
\textbf{\makecell{Lv.1 \\ Cluster}} & \textbf{\makecell{Lv.2 \\ Cluster}} & \textbf{\makecell{Lv.3 \\ Cluster}} &
\textbf{Label} & \textbf{Toxicity} & \textbf{Count} \\ 
\midrule

\multirow{2}{*}{\makecell{}} & \multirow{2}{*}{\makecell{11-78 \\ (Topic 78)}} & \multirow{2}{*}{\makecell{}} & \makecell{Immigration Support and Resources}  & \makecell{20.28} & \makecell{9,172} \\
\cmidrule(lr){4-6}
&  &  & \multicolumn{3}{l}{\makecell[{{l}}]{The topic highlights support for immigrants through resources, services, and \\ initiatives, promoting empowerment, integration, and a welcoming environment.}} \\
\midrule
\multirow{2}{*}{\makecell{}} & \multirow{2}{*}{\makecell{11-79 \\ (Topic 79)}} & \multirow{2}{*}{\makecell{}} & \makecell{Immigration and Gun Violence}  & \makecell{64.29} & \makecell{16,190} \\
\cmidrule(lr){4-6}
&  &  & \multicolumn{3}{l}{\makecell[{{l}}]{The topic highlights concerns over crimes by undocumented immigrants involving \\ gun violence, politicizing immigration policies and advocating stricter \\ controls.}} \\
\midrule
\multirow{2}{*}{\makecell{}} & \multirow{2}{*}{\makecell{11-80 \\ (Topic 80)}} & \multirow{2}{*}{\makecell{}} & \makecell{Illegal Immigration and Crime}  & \makecell{67.35} & \makecell{1,213} \\
\cmidrule(lr){4-6}
&  &  & \multicolumn{3}{l}{\makecell[{{l}}]{The topic highlights illegal immigration in the U.S., focusing on crimes by \\ undocumented immigrants and calls for stricter border control.}} \\
\midrule
\multirow{2}{*}{\makecell{}} & \multirow{2}{*}{\makecell{11-81 \\ (Topic 81)}} & \multirow{2}{*}{\makecell{}} & \makecell{Anti-Muslim Immigration Sentiment}  & \makecell{69.02} & \makecell{25,435} \\
\cmidrule(lr){4-6}
&  &  & \multicolumn{3}{l}{\makecell[{{l}}]{The topic highlights negative sentiments, conspiracy theories, and \\ discriminatory discourse against Muslim immigration, linking it to terrorism and \\ societal threats.}} \\
\midrule
\multirow{2}{*}{\makecell{}} & \multirow{2}{*}{\makecell{11-82 \\ (Topic 82)}} & \multirow{2}{*}{\makecell{}} & \makecell{Immigration and Crime Rates}  & \makecell{55.95} & \makecell{2,563} \\
\cmidrule(lr){4-6}
&  &  & \multicolumn{3}{l}{\makecell[{{l}}]{The topic discusses lower crime rates among immigrants, countering stereotypes \\ with studies and emphasizing their positive societal contributions.}} \\
\midrule
\multirow{2}{*}{\makecell{}} & \multirow{2}{*}{\makecell{11-83 \\ (Topic 83)}} & \multirow{2}{*}{\makecell{}} & \makecell{Immigration and Crime Association}  & \makecell{82.99} & \makecell{7,897} \\
\cmidrule(lr){4-6}
&  &  & \multicolumn{3}{l}{\makecell[{{l}}]{The topic explores the perceived link between immigration and crime, fueling \\ fears, misinformation, and calls for stricter border controls.}} \\
\midrule
\multirow{2}{*}{\makecell{}} & \multirow{2}{*}{\makecell{11-84 \\ (Topic 84)}} & \multirow{2}{*}{\makecell{}} & \makecell{Merit-Based Immigration Debate}  & \makecell{43.18} & \makecell{1,089} \\
\cmidrule(lr){4-6}
&  &  & \multicolumn{3}{l}{\makecell[{{l}}]{The topic discusses the debate on a merit-based U.S. immigration system, \\ addressing its benefits, criticisms, and related policy complexities.}} \\
\midrule
\multirow{2}{*}{\makecell{}} & \multirow{2}{*}{\makecell{11-85 \\ (Topic 85)}} & \multirow{2}{*}{\makecell{}} & \makecell{Immigration and Housing Crisis}  & \makecell{48.44} & \makecell{26,473} \\
\cmidrule(lr){4-6}
&  &  & \multicolumn{3}{l}{\makecell[{{l}}]{This topic explores the link between immigration and housing crises, \\ highlighting concerns about demand, affordability, and contributing factors.}} \\
\midrule
\multirow{2}{*}{\makecell{}} & \multirow{2}{*}{\makecell{11-86 \\ (Topic 86)}} & \multirow{2}{*}{\makecell{}} & \makecell{Blaming Immigrants for Social Issues}  & \makecell{55.00} & \makecell{4,362} \\
\cmidrule(lr){4-6}
&  &  & \multicolumn{3}{l}{\makecell[{{l}}]{This topic addresses the scapegoating of immigrants for social, economic, and \\ political issues, sparking polarized debates on immigration's impact.}} \\
\midrule
\multirow{2}{*}{\makecell{}} & \multirow{2}{*}{\makecell{11-87 \\ (Topic 87)}} & \multirow{2}{*}{\makecell{}} & \makecell{Immigrant Work Ethic}  & \makecell{49.73} & \makecell{1,586} \\
\cmidrule(lr){4-6}
&  &  & \multicolumn{3}{l}{\makecell[{{l}}]{The topic highlights immigrants' strong work ethic, economic contributions, and \\ challenges stereotypes, advocating for recognition of their societal value.}} \\
\midrule
\multirow{2}{*}{\makecell{}} & \multirow{2}{*}{\makecell{11-88 \\ (Topic 88)}} & \multirow{2}{*}{\makecell{}} & \makecell{Republican Rhetoric on Immigration}  & \makecell{52.67} & \makecell{1,179} \\
\cmidrule(lr){4-6}
&  &  & \multicolumn{3}{l}{\makecell[{{l}}]{The topic discusses claims that Republican "open borders" rhetoric fuels migrant \\ surges, highlighting political polarization and misinformation on immigration.}} \\
\midrule
\multirow{2}{*}{\makecell{}} & \multirow{2}{*}{\makecell{11-89 \\ (Topic 89)}} & \multirow{2}{*}{\makecell{}} & \makecell{Replacement Migration Conspiracy}  & \makecell{60.56} & \makecell{2,277} \\
\cmidrule(lr){4-6}
&  &  & \multicolumn{3}{l}{\makecell[{{l}}]{The topic explores "replacement migration" and its link to the "Great \\ Replacement" theory, highlighting fears of demographic and cultural shifts.}} \\
\midrule
\multirow{2}{*}{\makecell{}} & \multirow{2}{*}{\makecell{11-90 \\ (Topic 90)}} & \multirow{2}{*}{\makecell{}} & \makecell{Pro-Immigration Sentiment}  & \makecell{41.71} & \makecell{1,064} \\
\cmidrule(lr){4-6}
&  &  & \multicolumn{3}{l}{\makecell[{{l}}]{This topic highlights supportive tweets advocating for immigrant inclusion, \\ emphasizing their contributions, rights, and a more open immigration policy.}} \\

\bottomrule
\end{tabular}
\caption{Detailed Topic Discovery Results 8}
\label{table}
\end{table*}

\begin{table*}[ht]
\centering
\begin{tabular}{
  >{\centering\arraybackslash}m{1.2cm}
  >{\centering\arraybackslash}m{1.5cm}
  >{\centering\arraybackslash}m{1.2cm}
  >{\centering\arraybackslash}m{6.9cm}
  >{\centering\arraybackslash}m{1.5cm}
  >{\centering\arraybackslash}m{2.0cm}
}
\toprule
\textbf{\makecell{Lv.1 \\ Cluster}} & \textbf{\makecell{Lv.2 \\ Cluster}} & \textbf{\makecell{Lv.3 \\ Cluster}} &
\textbf{Label} & \textbf{Toxicity} & \textbf{Count} \\ 
\midrule

\multirow{2}{*}{\makecell{}} & \multirow{2}{*}{\makecell{11-91 \\ (Topic 91)}} & \multirow{2}{*}{\makecell{}} & \makecell{Migrant Child Trafficking Concerns}  & \makecell{68.60} & \makecell{11,884} \\
\cmidrule(lr){4-6}
&  &  & \multicolumn{3}{l}{\makecell[{{l}}]{The topic highlights concerns over migrant child trafficking, unaccompanied \\ minors, and criticism of the Biden Administration's immigration policies.}} \\
\midrule
\multirow{2}{*}{\makecell{}} & \multirow{2}{*}{\makecell{11-92 \\ (Topic 92)}} & \multirow{2}{*}{\makecell{}} & \makecell{Migrant Housing Crisis in NYC}  & \makecell{45.90} & \makecell{66,591} \\
\cmidrule(lr){4-6}
&  &  & \multicolumn{3}{l}{\makecell[{{l}}]{The topic highlights debates and tensions over migrant housing in NYC, involving \\ public facilities, safety concerns, and political responses.}} \\
\midrule
\multirow{2}{*}{\makecell{}} & \multirow{2}{*}{\makecell{}} & \multirow{2}{*}{\makecell{11-92-1}} & \makecell{Housing for Immigrants}  & \makecell{} & \makecell{6,516} \\
\cmidrule(lr){4-6}
&  &  & \multicolumn{3}{l}{\makecell[{{l}}]{The topic critiques U.S. immigration policies, suggesting housing migrants in \\ supporters' homes or properties, often using satire or sarcasm.}} \\
\midrule
\multirow{2}{*}{\makecell{}} & \multirow{2}{*}{\makecell{}} & \multirow{2}{*}{\makecell{11-92-2}} & \makecell{NYC Migrant Crisis Costs}  & \makecell{} & \makecell{2,487} \\
\cmidrule(lr){4-6}
&  &  & \multicolumn{3}{l}{\makecell[{{l}}]{The topic highlights New York City's \$12 billion migrant crisis cost, sparking \\ concerns over budgets, taxpayer strain, and federal assistance.}} \\
\midrule
\multirow{2}{*}{\makecell{}} & \multirow{2}{*}{\makecell{}} & \multirow{2}{*}{\makecell{11-92-3}} & \makecell{US Migrant Crisis Response}  & \makecell{} & \makecell{28,949} \\
\cmidrule(lr){4-6}
&  &  & \multicolumn{3}{l}{\makecell[{{l}}]{The topic covers the U.S. migrant crisis, focusing on challenges, political \\ responses, and social-economic implications, particularly in New York City.}} \\
\midrule
\multirow{2}{*}{\makecell{}} & \multirow{2}{*}{\makecell{11-93 \\ (Topic 93)}} & \multirow{2}{*}{\makecell{}} & \makecell{US Immigration and Border Crisis}  & \makecell{38.42} & \makecell{8,005} \\
\cmidrule(lr){4-6}
&  &  & \multicolumn{3}{l}{\makecell[{{l}}]{The topic covers U.S. immigration challenges, migrant surges, root causes, \\ policy impacts, and debates on border control and compassion.}} \\
\midrule
\multirow{2}{*}{\makecell{}} & \multirow{2}{*}{\makecell{}} & \multirow{2}{*}{\makecell{11-93-1}} & \makecell{U.S.-Mexico Border Crisis}  & \makecell{} & \makecell{1,864} \\
\cmidrule(lr){4-6}
&  &  & \multicolumn{3}{l}{\makecell[{{l}}]{This topic focuses on the U.S.-Mexico border crisis, migrant surges, border \\ security, immigration policies, and related political and humanitarian issues.}} \\
\midrule
\multirow{2}{*}{\makecell{}} & \multirow{2}{*}{\makecell{}} & \multirow{2}{*}{\makecell{11-93-2}} & \makecell{Darien Gap Migration Crisis}  & \makecell{} & \makecell{1,834} \\
\cmidrule(lr){4-6}
&  &  & \multicolumn{3}{l}{\makecell[{{l}}]{The topic covers the migrant surge through the dangerous Darien Gap, \\ highlighting hardships, deaths, and debates on immigration policies.}} \\
\midrule
\multirow{2}{*}{\makecell{}} & \multirow{2}{*}{\makecell{11-94 \\ (Topic 94)}} & \multirow{2}{*}{\makecell{}} & \makecell{Trump's Anti-Immigrant Rhetoric}  & \makecell{77.64} & \makecell{1,952} \\
\cmidrule(lr){4-6}
&  &  & \multicolumn{3}{l}{\makecell[{{l}}]{The topic highlights backlash against Donald Trump's anti-immigrant rhetoric, \\ including his ``poisoning the blood'' comment likened to Nazi ideology.}} \\
\midrule
\multirow{2}{*}{\makecell{}} & \multirow{2}{*}{\makecell{11-95 \\ (Topic 95)}} & \multirow{2}{*}{\makecell{}} & \makecell{Immigration and Social Unrest}  & \makecell{63.02} & \makecell{4,115} \\
\cmidrule(lr){4-6}
&  &  & \multicolumn{3}{l}{\makecell[{{l}}]{The topic explores immigration's link to social unrest, focusing on riots, \\ perceptions, policies, media influence, and public opinion.}} \\
\midrule
\multirow{2}{*}{\makecell{}} & \multirow{2}{*}{\makecell{11-96 \\ (Topic 96)}} & \multirow{2}{*}{\makecell{}} & \makecell{Legal vs Illegal Immigration}  & \makecell{47.28} & \makecell{44,491} \\
\cmidrule(lr){4-6}
&  &  & \multicolumn{3}{l}{\makecell[{{l}}]{The topic highlights debates on legal vs. illegal immigration in the U.S., \\ emphasizing law, politics, and societal impacts.}} \\
\midrule
\multirow{2}{*}{\makecell{}} & \multirow{2}{*}{\makecell{}} & \multirow{2}{*}{\makecell{11-96-1}} & \makecell{Legal vs Illegal Immigration}  & \makecell{} & \makecell{1,763} \\
\cmidrule(lr){4-6}
&  &  & \multicolumn{3}{l}{\makecell[{{l}}]{This topic highlights the contentious debate over distinguishing legal and \\ illegal immigration in U.S. immigration policy discussions.}} \\
\midrule
\multirow{2}{*}{\makecell{}} & \multirow{2}{*}{\makecell{}} & \multirow{2}{*}{\makecell{11-96-2}} & \makecell{Legal Immigration Debate}  & \makecell{} & \makecell{13,864} \\
\cmidrule(lr){4-6}
&  &  & \multicolumn{3}{l}{\makecell[{{l}}]{The topic discusses support for legal immigration, criticism of inefficiencies, \\ and calls for clearer policies and political accountability.}} \\

\bottomrule
\end{tabular}
\caption{Detailed Topic Discovery Results 9}
\label{table}
\end{table*}

\begin{table*}[ht]
\centering
\begin{tabular}{
  >{\centering\arraybackslash}m{1.2cm}
  >{\centering\arraybackslash}m{1.5cm}
  >{\centering\arraybackslash}m{1.2cm}
  >{\centering\arraybackslash}m{6.9cm}
  >{\centering\arraybackslash}m{1.5cm}
  >{\centering\arraybackslash}m{2.0cm}
}
\toprule
\textbf{\makecell{Lv.1 \\ Cluster}} & \textbf{\makecell{Lv.2 \\ Cluster}} & \textbf{\makecell{Lv.3 \\ Cluster}} &
\textbf{Label} & \textbf{Toxicity} & \textbf{Count} \\ 
\midrule

\multirow{2}{*}{\makecell{}} & \multirow{2}{*}{\makecell{11-97 \\ (Topic 97)}} & \multirow{2}{*}{\makecell{}} & \makecell{African Immigration and Colonialism}  & \makecell{60.54} & \makecell{1,485} \\
\cmidrule(lr){4-6}
&  &  & \multicolumn{3}{l}{\makecell[{{l}}]{The topic discusses African migration to the West, linking it to colonialism, \\ Western responsibility, and African governance challenges.}} \\
\midrule
\multirow{2}{*}{\makecell{}} & \multirow{2}{*}{\makecell{11-98 \\ (Topic 98)}} & \multirow{2}{*}{\makecell{}} & \makecell{Western Intervention and Migration}  & \makecell{58.25} & \makecell{9,750} \\
\cmidrule(lr){4-6}
&  &  & \multicolumn{3}{l}{\makecell[{{l}}]{Western military interventions and regime changes have caused mass migration, \\ refugee crises, and tensions in Europe, sparking widespread debate.}} \\
\midrule
\multirow{2}{*}{\makecell{}} & \multirow{2}{*}{\makecell{11-99 \\ (Topic 99)}} & \multirow{2}{*}{\makecell{}} & \makecell{Anti-Immigration Conspiracy Theories}  & \makecell{64.56} & \makecell{7,207} \\
\cmidrule(lr){4-6}
&  &  & \multicolumn{3}{l}{\makecell[{{l}}]{The topic involves tweets promoting conspiracy theories linking immigration to \\ societal decline, global elite control, and the erosion of national identity.}} \\
\midrule
\multirow{2}{*}{\makecell{}} & \multirow{2}{*}{\makecell{11-100 \\ (Topic 100)}} & \multirow{2}{*}{\makecell{}} & \makecell{Controlled Immigration Debate}  & \makecell{46.28} & \makecell{1,095} \\
\cmidrule(lr){4-6}
&  &  & \multicolumn{3}{l}{\makecell[{{l}}]{The topic discusses debates on regulated immigration policies, balancing \\ economic, security, and humanitarian concerns, with varying opinions on stricter \\ controls.}} \\
\midrule
\multirow{2}{*}{\makecell{}} & \multirow{2}{*}{\makecell{11-101 \\ (Topic 101)}} & \multirow{2}{*}{\makecell{}} & \makecell{Immigration and Social Justice Issues}  & \makecell{69.61} & \makecell{11,215} \\
\cmidrule(lr){4-6}
&  &  & \multicolumn{3}{l}{\makecell[{{l}}]{This topic covers U.S. immigration issues, social justice, LGBTQ+ and racial \\ minority rights, and polarized views on related policies and ideologies.}} \\
\midrule
\multirow{2}{*}{\makecell{}} & \multirow{2}{*}{\makecell{11-102 \\ (Topic 102)}} & \multirow{2}{*}{\makecell{}} & \makecell{Reducing Immigration Levels}  & \makecell{47.73} & \makecell{2,961} \\
\cmidrule(lr){4-6}
&  &  & \multicolumn{3}{l}{\makecell[{{l}}]{The topic focuses on reducing immigration levels, particularly low-skilled \\ immigrants, with criticism of current policies and government inaction.}} \\
\midrule
\multirow{2}{*}{\makecell{}} & \multirow{2}{*}{\makecell{11-103 \\ (Topic 103)}} & \multirow{2}{*}{\makecell{}} & \makecell{Immigration Numbers and Policies}  & \makecell{42.64} & \makecell{1,233} \\
\cmidrule(lr){4-6}
&  &  & \multicolumn{3}{l}{\makecell[{{l}}]{This topic discusses ideal immigration levels, policy impacts, and economic, \\ social, and environmental implications of varying immigration rates.}} \\
\midrule
\multirow{2}{*}{\makecell{}} & \multirow{2}{*}{\makecell{11-104 \\ (Topic 104)}} & \multirow{2}{*}{\makecell{}} & \makecell{Immigration and Cultural Assimilation}  & \makecell{54.63} & \makecell{8,461} \\
\cmidrule(lr){4-6}
&  &  & \multicolumn{3}{l}{\makecell[{{l}}]{The topic discusses immigration and cultural assimilation, focusing on concerns \\ about immigrants adapting to the host country's culture.}} \\
\midrule
\multirow{2}{*}{\makecell{}} & \multirow{2}{*}{\makecell{11-105 \\ (Topic 105)}} & \multirow{2}{*}{\makecell{}} & \makecell{Immigration and Colonialism Comparison}  & \makecell{57.73} & \makecell{1,213} \\
\cmidrule(lr){4-6}
&  &  & \multicolumn{3}{l}{\makecell[{{l}}]{The topic explores the conflation of immigration and colonialism, critiquing \\ Western nations' immigration policies in light of their colonial histories.}} \\
\midrule
\multirow{2}{*}{\makecell{}} & \multirow{2}{*}{\makecell{11-106 \\ (Topic 106)}} & \multirow{2}{*}{\makecell{}} & \makecell{Immigration as Invasion}  & \makecell{62.92} & \makecell{9,407} \\
\cmidrule(lr){4-6}
&  &  & \multicolumn{3}{l}{\makecell[{{l}}]{This topic frames U.S. immigration as an ``invasion,'' using inflammatory language \\ to depict migrants as threats to security and identity.}} \\
\midrule
\multirow{2}{*}{\makecell{}} & \multirow{2}{*}{\makecell{11-107 \\ (Topic 107)}} & \multirow{2}{*}{\makecell{}} & \makecell{Anti-Immigration Sentiment}  & \makecell{62.12} & \makecell{1,493} \\
\cmidrule(lr){4-6}
&  &  & \multicolumn{3}{l}{\makecell[{{l}}]{The topic highlights anti-immigration sentiment, citing cultural erosion, \\ national identity loss, and advocating stricter border controls against mass \\ immigration.}} \\
\midrule
\multirow{2}{*}{\makecell{}} & \multirow{2}{*}{\makecell{11-108 \\ (Topic 108)}} & \multirow{2}{*}{\makecell{}} & \makecell{Opposition to Mass Immigration}  & \makecell{50.42} & \makecell{6,211} \\
\cmidrule(lr){4-6}
&  &  & \multicolumn{3}{l}{\makecell[{{l}}]{The topic highlights anti-mass immigration sentiment, criticizing governments \\ and advocating stricter border controls due to cultural, economic, and security \\ concerns.}} \\
\midrule
\multirow{2}{*}{\makecell{}} & \multirow{2}{*}{\makecell{11-109 \\ (Topic 109)}} & \multirow{2}{*}{\makecell{}} & \makecell{Immigration and Veteran Welfare}  & \makecell{62.49} & \makecell{20,100} \\
\cmidrule(lr){4-6}
&  &  & \multicolumn{3}{l}{\makecell[{{l}}]{The topic highlights a debate over perceived prioritization of illegal \\ immigrants' welfare over U.S. veterans' support, sparking political division.}} \\

\bottomrule
\end{tabular}
\caption{Detailed Topic Discovery Results 10}
\label{table}
\end{table*}

\begin{table*}[ht]
\centering
\begin{tabular}{
  >{\centering\arraybackslash}m{1.2cm}
  >{\centering\arraybackslash}m{1.5cm}
  >{\centering\arraybackslash}m{1.2cm}
  >{\centering\arraybackslash}m{6.9cm}
  >{\centering\arraybackslash}m{1.5cm}
  >{\centering\arraybackslash}m{2.0cm}
}
\toprule
\textbf{\makecell{Lv.1 \\ Cluster}} & \textbf{\makecell{Lv.2 \\ Cluster}} & \textbf{\makecell{Lv.3 \\ Cluster}} &
\textbf{Label} & \textbf{Toxicity} & \textbf{Count} \\ 
\midrule

\multirow{2}{*}{\makecell{}} & \multirow{2}{*}{\makecell{11-110 \\ (Topic 110)}} & \multirow{2}{*}{\makecell{}} & \makecell{Anti-Immigration Sentiment}  & \makecell{55.67} & \makecell{1,641} \\
\cmidrule(lr){4-6}
&  &  & \multicolumn{3}{l}{\makecell[{{l}}]{The topic highlights tweets opposing U.S. immigration, emphasizing negative \\ impacts, urgent language, and calls for halting immigration entirely.}} \\
\midrule
\multirow{2}{*}{\makecell{}} & \multirow{2}{*}{\makecell{11-111 \\ (Topic 111)}} & \multirow{2}{*}{\makecell{}} & \makecell{Uncontrolled Immigration Concerns}  & \makecell{51.37} & \makecell{2,791} \\
\cmidrule(lr){4-6}
&  &  & \multicolumn{3}{l}{\makecell[{{l}}]{The topic highlights concerns about unchecked immigration in the U.S., focusing \\ on social, economic, and cultural impacts with critical sentiments.}} \\
\midrule
\multirow{2}{*}{\makecell{}} & \multirow{2}{*}{\makecell{11-112 \\ (Topic 112)}} & \multirow{2}{*}{\makecell{}} & \makecell{E-Verify and Immigration Reform}  & \makecell{51.75} & \makecell{1,015} \\
\cmidrule(lr){4-6}
&  &  & \multicolumn{3}{l}{\makecell[{{l}}]{This topic discusses E-Verify's role in immigration reform, debating its \\ effectiveness, implications, and connection to broader immigration issues.}} \\
\midrule
\multirow{2}{*}{\makecell{}} & \multirow{2}{*}{\makecell{11-113 \\ (Topic 113)}} & \multirow{2}{*}{\makecell{}} & \makecell{Skilled Immigration Debate}  & \makecell{42.55} & \makecell{1,155} \\
\cmidrule(lr){4-6}
&  &  & \multicolumn{3}{l}{\makecell[{{l}}]{The topic discusses debates on skilled immigration in the U.S., highlighting \\ benefits, drawbacks, and calls for a points-based system.}} \\
\midrule
\multirow{2}{*}{\makecell{}} & \multirow{2}{*}{\makecell{11-114 \\ (Topic 114)}} & \multirow{2}{*}{\makecell{}} & \makecell{Immigration and Demographic Shifts}  & \makecell{51.32} & \makecell{5,649} \\
\cmidrule(lr){4-6}
&  &  & \multicolumn{3}{l}{\makecell[{{l}}]{This topic discusses immigration as a solution to U.S. population decline, \\ addressing low birth rates and economic implications.}} \\
\midrule
\multirow{2}{*}{\makecell{}} & \multirow{2}{*}{\makecell{11-115 \\ (Topic 115)}} & \multirow{2}{*}{\makecell{}} & \makecell{Labor Shortage and Immigration}  & \makecell{44.68} & \makecell{2,255} \\
\cmidrule(lr){4-6}
&  &  & \multicolumn{3}{l}{\makecell[{{l}}]{The U.S. faces labor shortages, and immigration is proposed as a solution to \\ fill jobs, boost the economy, and alleviate crises.}} \\
\midrule
\multirow{2}{*}{\makecell{}} & \multirow{2}{*}{\makecell{11-116 \\ (Topic 116)}} & \multirow{2}{*}{\makecell{}} & \makecell{Economic Benefits of Immigration}  & \makecell{39.58} & \makecell{1,099} \\
\cmidrule(lr){4-6}
&  &  & \multicolumn{3}{l}{\makecell[{{l}}]{This topic highlights immigration's positive economic impact, emphasizing labor \\ gap filling, productivity, growth, and countering negative stereotypes.}} \\
\midrule
\multirow{2}{*}{\makecell{}} & \multirow{2}{*}{\makecell{11-117 \\ (Topic 117)}} & \multirow{2}{*}{\makecell{}} & \makecell{Hiring of Illegal Immigrants}  & \makecell{56.68} & \makecell{14,658} \\
\cmidrule(lr){4-6}
&  &  & \multicolumn{3}{l}{\makecell[{{l}}]{The topic discusses companies hiring illegal immigrants, legal implications, \\ government enforcement, and public debate over U.S. immigration policies.}} \\
\midrule
\multirow{2}{*}{\makecell{}} & \multirow{2}{*}{\makecell{11-118 \\ (Topic 118)}} & \multirow{2}{*}{\makecell{}} & \makecell{Elon Musk's Immigration Hypocrisy}  & \makecell{48.20} & \makecell{1,085} \\
\cmidrule(lr){4-6}
&  &  & \multicolumn{3}{l}{\makecell[{{l}}]{The topic highlights accusations of hypocrisy against Elon Musk, a South African \\ immigrant, for criticizing U.S. immigration policies despite benefiting from \\ immigration.}} \\
\midrule
\multirow{2}{*}{\makecell{}} & \multirow{2}{*}{\makecell{11-119 \\ (Topic 119)}} & \multirow{2}{*}{\makecell{}} & \makecell{U.S. Border Immigration Concerns}  & \makecell{65.13} & \makecell{16,211} \\
\cmidrule(lr){4-6}
&  &  & \multicolumn{3}{l}{\makecell[{{l}}]{The topic highlights concerns over U.S. southern border issues, illegal \\ immigration, drug trafficking, and calls for stricter border control.}} \\
\midrule
\multirow{2}{*}{\makecell{}} & \multirow{2}{*}{\makecell{11-120 \\ (Topic 120)}} & \multirow{2}{*}{\makecell{}} & \makecell{DeSantis' Immigration Policies in Florida}  & \makecell{51.96} & \makecell{7,691} \\
\cmidrule(lr){4-6}
&  &  & \multicolumn{3}{l}{\makecell[{{l}}]{The topic discusses criticism of Florida Governor Ron DeSantis' immigration \\ policies, highlighting their potential harm to the state's economy and \\ industries.}} \\
\midrule
\multirow{2}{*}{\makecell{}} & \multirow{2}{*}{\makecell{11-121 \\ (Topic 121)}} & \multirow{2}{*}{\makecell{}} & \makecell{Immigration and Farm Labor}  & \makecell{54.98} & \makecell{1,683} \\
\cmidrule(lr){4-6}
&  &  & \multicolumn{3}{l}{\makecell[{{l}}]{The topic examines U.S. immigration's impact on agriculture, highlighting \\ migrant workers' essential role, economic effects, and policy contradictions.}} \\
\midrule
\multirow{2}{*}{\makecell{}} & \multirow{2}{*}{\makecell{11-122 \\ (Topic 122)}} & \multirow{2}{*}{\makecell{}} & \makecell{Immigration and Labor Economics}  & \makecell{55.42} & \makecell{9,611} \\
\cmidrule(lr){4-6}
&  &  & \multicolumn{3}{l}{\makecell[{{l}}]{The topic examines immigration's impact on wages, job availability, and labor \\ economics, highlighting debates over policy, inequality, and corporate roles.}} \\

\bottomrule
\end{tabular}
\caption{Detailed Topic Discovery Results 11}
\label{table}
\end{table*}

\begin{table*}[ht]
\centering
\begin{tabular}{
  >{\centering\arraybackslash}m{1.2cm}
  >{\centering\arraybackslash}m{1.5cm}
  >{\centering\arraybackslash}m{1.2cm}
  >{\centering\arraybackslash}m{6.9cm}
  >{\centering\arraybackslash}m{1.5cm}
  >{\centering\arraybackslash}m{2.0cm}
}
\toprule
\textbf{\makecell{Lv.1 \\ Cluster}} & \textbf{\makecell{Lv.2 \\ Cluster}} & \textbf{\makecell{Lv.3 \\ Cluster}} &
\textbf{Label} & \textbf{Toxicity} & \textbf{Count} \\ 
\midrule

\multirow{2}{*}{\makecell{}} & \multirow{2}{*}{\makecell{11-123 \\ (Topic 123)}} & \multirow{2}{*}{\makecell{}} & \makecell{Immigration and Job Competition}  & \makecell{62.69} & \makecell{17,645} \\
\cmidrule(lr){4-6}
&  &  & \multicolumn{3}{l}{\makecell[{{l}}]{The topic discusses concerns about illegal immigration's impact on U.S. jobs, \\ wages, labor laws, and calls for stricter regulations.}} \\
\midrule
\multirow{2}{*}{\makecell{}} & \multirow{2}{*}{\makecell{11-124 \\ (Topic 124)}} & \multirow{2}{*}{\makecell{}} & \makecell{Challenging "Illegal" Immigration Labels}  & \makecell{48.08} & \makecell{3,757} \\
\cmidrule(lr){4-6}
&  &  & \multicolumn{3}{l}{\makecell[{{l}}]{The topic debates the term "illegal" for immigrants, advocating for inclusive \\ language and highlighting migrants' positive societal contributions.}} \\
\midrule
\multirow{2}{*}{\makecell{}} & \multirow{2}{*}{\makecell{11-125 \\ (Topic 125)}} & \multirow{2}{*}{\makecell{}} & \makecell{Colony Ridge Immigration Controversy}  & \makecell{59.88} & \makecell{3,000} \\
\cmidrule(lr){4-6}
&  &  & \multicolumn{3}{l}{\makecell[{{l}}]{The controversy over Texas's Colony Ridge housing development allegedly housing \\ illegal immigrants sparks concerns about crime, governance, and inaction.}} \\
\midrule
\multirow{2}{*}{\makecell{}} & \multirow{2}{*}{\makecell{11-126 \\ (Topic 126)}} & \multirow{2}{*}{\makecell{}} & \makecell{U.S. Border Immigration Crisis}  & \makecell{57.29} & \makecell{4,971} \\
\cmidrule(lr){4-6}
&  &  & \multicolumn{3}{l}{\makecell[{{l}}]{The topic highlights debates on U.S.-Mexico border issues, focusing on illegal \\ immigration, border security, and polarized views on policy approaches.}} \\
\midrule
\multirow{2}{*}{\makecell{}} & \multirow{2}{*}{\makecell{11-127 \\ (Topic 127)}} & \multirow{2}{*}{\makecell{}} & \makecell{Anti-Immigration Sentiment}  & \makecell{70.58} & \makecell{9,742} \\
\cmidrule(lr){4-6}
&  &  & \multicolumn{3}{l}{\makecell[{{l}}]{This topic covers hostile tweets on U.S. illegal immigration, using derogatory \\ language, criticizing policies, and promoting divisive rhetoric.}} \\
\midrule
\multirow{2}{*}{\makecell{}} & \multirow{2}{*}{\makecell{11-128 \\ (Topic 128)}} & \multirow{2}{*}{\makecell{}} & \makecell{Immigrant Experiences and Stories}  & \makecell{26.71} & \makecell{7,196} \\
\cmidrule(lr){4-6}
&  &  & \multicolumn{3}{l}{\makecell[{{l}}]{The topic highlights immigrant experiences in media, emphasizing authentic \\ representation in films, books, art, and cultural storytelling.}} \\
\midrule
\multirow{2}{*}{\makecell{}} & \multirow{2}{*}{\makecell{11-129 \\ (Topic 129)}} & \multirow{2}{*}{\makecell{}} & \makecell{Immigrant Cultural Exchange}  & \makecell{39.87} & \makecell{2,150} \\
\cmidrule(lr){4-6}
&  &  & \multicolumn{3}{l}{\makecell[{{l}}]{The topic highlights immigrants enriching local culture through unique culinary \\ contributions, fusion dishes, and immigrant-owned restaurants, fostering \\ diversity.}} \\
\midrule
\multirow{2}{*}{\makecell{}} & \multirow{2}{*}{\makecell{11-130 \\ (Topic 130)}} & \multirow{2}{*}{\makecell{}} & \makecell{U.S. Border Control Debate}  & \makecell{61.95} & \makecell{10,098} \\
\cmidrule(lr){4-6}
&  &  & \multicolumn{3}{l}{\makecell[{{l}}]{The topic highlights concerns over U.S. immigration, emphasizing stricter \\ southern border control, national security, and the economic impact of illegal \\ crossings.}} \\
\midrule
\multirow{2}{*}{\makecell{}} & \multirow{2}{*}{\makecell{11-131 \\ (Topic 131)}} & \multirow{2}{*}{\makecell{}} & \makecell{Opposition to Immigrant Housing}  & \makecell{61.78} & \makecell{6,356} \\
\cmidrule(lr){4-6}
&  &  & \multicolumn{3}{l}{\makecell[{{l}}]{The topic highlights frustration over housing resources for illegal immigrants, \\ perceived as prioritized over homeless Americans, sparking criticism of \\ immigration policies.}} \\
\midrule
\multirow{2}{*}{\makecell{}} & \multirow{2}{*}{\makecell{11-132 \\ (Topic 132)}} & \multirow{2}{*}{\makecell{}} & \makecell{Anti-Immigration White Nationalism}  & \makecell{76.81} & \makecell{2,298} \\
\cmidrule(lr){4-6}
&  &  & \multicolumn{3}{l}{\makecell[{{l}}]{The topic involves tweets with anti-immigration sentiments, white nationalist \\ ideologies, and alarmist claims about immigration eroding white majority \\ populations.}} \\
\midrule
\multirow{2}{*}{\makecell{}} & \multirow{2}{*}{\makecell{11-133 \\ (Topic 133)}} & \multirow{2}{*}{\makecell{}} & \makecell{California Immigration and Crime Issues}  & \makecell{67.00} & \makecell{2,828} \\
\cmidrule(lr){4-6}
&  &  & \multicolumn{3}{l}{\makecell[{{l}}]{Criticism of California's immigration, crime, and social issues under Governor \\ Newsom, highlighting homelessness, high taxes, and illegal immigration concerns.}} \\
\midrule
\multirow{2}{*}{\makecell{}} & \multirow{2}{*}{\makecell{11-134 \\ (Topic 134)}} & \multirow{2}{*}{\makecell{}} & \makecell{Racial Bias in Immigration}  & \makecell{66.61} & \makecell{4,648} \\
\cmidrule(lr){4-6}
&  &  & \multicolumn{3}{l}{\makecell[{{l}}]{This topic examines racial bias in U.S. immigration policies, highlighting \\ disparities in treatment and perception based on skin color.}} \\
\midrule
\multirow{2}{*}{\makecell{}} & \multirow{2}{*}{\makecell{11-135 \\ (Topic 135)}} & \multirow{2}{*}{\makecell{}} & \makecell{Black American vs Immigrant Dynamics}  & \makecell{68.28} & \makecell{18,804} \\
\cmidrule(lr){4-6}
&  &  & \multicolumn{3}{l}{\makecell[{{l}}]{The topic explores tensions between Black Americans and black immigrants, \\ focusing on identity, culture, societal treatment, and shared complexities.}} \\

\bottomrule
\end{tabular}
\caption{Detailed Topic Discovery Results 12}
\label{table}
\end{table*}

\begin{table*}[ht]
\centering
\begin{tabular}{
  >{\centering\arraybackslash}m{1.2cm}
  >{\centering\arraybackslash}m{1.5cm}
  >{\centering\arraybackslash}m{1.2cm}
  >{\centering\arraybackslash}m{6.9cm}
  >{\centering\arraybackslash}m{1.5cm}
  >{\centering\arraybackslash}m{2.0cm}
}
\toprule
\textbf{\makecell{Lv.1 \\ Cluster}} & \textbf{\makecell{Lv.2 \\ Cluster}} & \textbf{\makecell{Lv.3 \\ Cluster}} &
\textbf{Label} & \textbf{Toxicity} & \textbf{Count} \\ 
\midrule

\multirow{2}{*}{\makecell{}} & \multirow{2}{*}{\makecell{11-136 \\ (Topic 136)}} & \multirow{2}{*}{\makecell{}} & \makecell{Busing of Migrants by Texas Governor}  & \makecell{47.59} & \makecell{7,050} \\
\cmidrule(lr){4-6}
&  &  & \multicolumn{3}{l}{\makecell[{{l}}]{The topic covers Texas Governor Greg Abbott's controversial policy of busing \\ migrants to Democratic-leaning areas, sparking political and humanitarian \\ debates.}} \\
\midrule
\multirow{2}{*}{\makecell{}} & \multirow{2}{*}{\makecell{11-137 \\ (Topic 137)}} & \multirow{2}{*}{\makecell{}} & \makecell{Illegal Immigration as Crime}  & \makecell{65.62} & \makecell{2,334} \\
\cmidrule(lr){4-6}
&  &  & \multicolumn{3}{l}{\makecell[{{l}}]{This topic focuses on framing illegal immigration as criminal, advocating strict \\ enforcement, deportation, and criticizing lenient immigration policies.}} \\
\midrule
\multirow{2}{*}{\makecell{}} & \multirow{2}{*}{\makecell{11-138 \\ (Topic 138)}} & \multirow{2}{*}{\makecell{}} & \makecell{Immigration and National Security}  & \makecell{68.13} & \makecell{13,762} \\
\cmidrule(lr){4-6}
&  &  & \multicolumn{3}{l}{\makecell[{{l}}]{The topic highlights concerns over U.S. immigration, focusing on national \\ security threats, illegal crossings, and criticism of border policies.}} \\
\midrule
\multirow{2}{*}{\makecell{}} & \multirow{2}{*}{\makecell{11-139 \\ (Topic 139)}} & \multirow{2}{*}{\makecell{}} & \makecell{Sanctuary Cities and Immigration}  & \makecell{59.09} & \makecell{21,501} \\
\cmidrule(lr){4-6}
&  &  & \multicolumn{3}{l}{\makecell[{{l}}]{The topic discusses the polarized debate on transporting illegal immigrants to \\ sanctuary cities, highlighting political controversy and perceived hypocrisy.}} \\
\midrule
\multirow{2}{*}{\makecell{}} & \multirow{2}{*}{\makecell{11-140 \\ (Topic 140)}} & \multirow{2}{*}{\makecell{}} & \makecell{Criticism of Immigration Hypocrisy}  & \makecell{63.33} & \makecell{6,869} \\
\cmidrule(lr){4-6}
&  &  & \multicolumn{3}{l}{\makecell[{{l}}]{The topic critiques perceived hypocrisy of immigration supporters, questioning \\ their willingness to personally house illegal immigrants, often sarcastically.}} \\
\midrule
\multirow{2}{*}{\makecell{}} & \multirow{2}{*}{\makecell{11-141 \\ (Topic 141)}} & \multirow{2}{*}{\makecell{}} & \makecell{Opposition to Immigration Spending}  & \makecell{60.57} & \makecell{1,376} \\
\cmidrule(lr){4-6}
&  &  & \multicolumn{3}{l}{\makecell[{{l}}]{Tweets criticize New York City’s spending on illegal immigrants, arguing it \\ neglects legal residents' needs, infrastructure, and essential services.}} \\
\midrule
\multirow{2}{*}{\makecell{}} & \multirow{2}{*}{\makecell{11-142 \\ (Topic 142)}} & \multirow{2}{*}{\makecell{}} & \makecell{Opposition to US Immigration Policies}  & \makecell{62.01} & \makecell{9,516} \\
\cmidrule(lr){4-6}
&  &  & \multicolumn{3}{l}{\makecell[{{l}}]{The topic highlights opposition to US immigration policies, criticizing \\ government inaction and prioritization of illegal immigrants over citizens.}} \\
\midrule
\multirow{2}{*}{\makecell{}} & \multirow{2}{*}{\makecell{11-143 \\ (Topic 143)}} & \multirow{2}{*}{\makecell{}} & \makecell{U.S. Immigration Border Crossings}  & \makecell{47.83} & \makecell{2,738} \\
\cmidrule(lr){4-6}
&  &  & \multicolumn{3}{l}{\makecell[{{l}}]{The topic covers social media debates on U.S. illegal border crossings, \\ immigration policy effectiveness, and partisan views on migrant issues.}} \\
\midrule
\multirow{2}{*}{\makecell{}} & \multirow{2}{*}{\makecell{11-144 \\ (Topic 144)}} & \multirow{2}{*}{\makecell{}} & \makecell{Illegal Immigration Concerns}  & \makecell{55.28} & \makecell{1,889} \\
\cmidrule(lr){4-6}
&  &  & \multicolumn{3}{l}{\makecell[{{l}}]{The topic highlights frustrations over illegal immigration in the U.S., \\ emphasizing law enforcement, national security, and criticism of enabling \\ policies.}} \\
\midrule
\multirow{2}{*}{\makecell{}} & \multirow{2}{*}{\makecell{11-145 \\ (Topic 145)}} & \multirow{2}{*}{\makecell{}} & \makecell{Immigration and Illegal Aliens}  & \makecell{61.30} & \makecell{1,987} \\
\cmidrule(lr){4-6}
&  &  & \multicolumn{3}{l}{\makecell[{{l}}]{The topic focuses on U.S. immigration issues, particularly illegal immigration, \\ highlighting debates, concerns, and divisive sentiments on related policies.}} \\
\midrule
\multirow{2}{*}{\makecell{}} & \multirow{2}{*}{\makecell{11-146 \\ (Topic 146)}} & \multirow{2}{*}{\makecell{}} & \makecell{Illegal Immigration Debate}  & \makecell{64.50} & \makecell{10,766} \\
\cmidrule(lr){4-6}
&  &  & \multicolumn{3}{l}{\makecell[{{l}}]{The topic addresses the polarized debate over terminology for undocumented \\ immigrants in the U.S., tied to border security and societal impact.}} \\
\midrule
\multirow{2}{*}{\makecell{}} & \multirow{2}{*}{\makecell{11-147 \\ (Topic 147)}} & \multirow{2}{*}{\makecell{}} & \makecell{U.S. Immigration Concerns}  & \makecell{61.67} & \makecell{43,217} \\
\cmidrule(lr){4-6}
&  &  & \multicolumn{3}{l}{\makecell[{{l}}]{The topic highlights criticisms of U.S. immigration policies, focusing on \\ illegal immigration, border control, and national security concerns.}} \\

\bottomrule
\end{tabular}
\caption{Detailed Topic Discovery Results 13}
\label{table}
\end{table*}

\begin{table*}[ht]
\centering
\begin{tabular}{
  >{\centering\arraybackslash}m{1.2cm}
  >{\centering\arraybackslash}m{1.5cm}
  >{\centering\arraybackslash}m{1.2cm}
  >{\centering\arraybackslash}m{6.9cm}
  >{\centering\arraybackslash}m{1.5cm}
  >{\centering\arraybackslash}m{2.0cm}
}
\toprule
\textbf{\makecell{Lv.1 \\ Cluster}} & \textbf{\makecell{Lv.2 \\ Cluster}} & \textbf{\makecell{Lv.3 \\ Cluster}} &
\textbf{Label} & \textbf{Toxicity} & \textbf{Count} \\ 
\midrule

\multirow{2}{*}{\makecell{}} & \multirow{2}{*}{\makecell{11-148 \\ (Topic 148)}} & \multirow{2}{*}{\makecell{}} & \makecell{Anti-Immigration Sentiment and Treason Accusations}  & \makecell{71.79} & \makecell{1,668} \\
\cmidrule(lr){4-6}
&  &  & \multicolumn{3}{l}{\makecell[{{l}}]{The topic highlights strong criticism of U.S. immigration policies, accusing the \\ Biden administration of enabling illegal immigration and undermining \\ sovereignty.}} \\
\midrule

\multirow{2}{*}{\makecell{}} & \multirow{2}{*}{\makecell{11-149 \\ (Topic 149)}} & \multirow{2}{*}{\makecell{}} & \makecell{Stopping Illegal US Immigration}  & \makecell{50.79} & \makecell{2,049} \\
\cmidrule(lr){4-6}
&  &  & \multicolumn{3}{l}{\makecell[{{l}}]{This topic focuses on reducing illegal immigration to the U.S., emphasizing \\ stricter border controls, national security, and economic concerns.}} \\
\midrule
\multirow{2}{*}{\makecell{}} & \multirow{2}{*}{\makecell{11-150 \\ (Topic 150)}} & \multirow{2}{*}{\makecell{}} & \makecell{Social Security for Illegal Immigrants}  & \makecell{63.71} & \makecell{4,014} \\
\cmidrule(lr){4-6}
&  &  & \multicolumn{3}{l}{\makecell[{{l}}]{Twitter users express outrage over claims that illegal immigrants receive Social \\ Security funds, surpassing payments to legal citizens and retirees.}} \\
\midrule
\multirow{2}{*}{\makecell{}} & \multirow{2}{*}{\makecell{11-151 \\ (Topic 151)}} & \multirow{2}{*}{\makecell{}} & \makecell{Healthcare for Illegal Immigrants}  & \makecell{62.69} & \makecell{2,492} \\
\cmidrule(lr){4-6}
&  &  & \multicolumn{3}{l}{\makecell[{{l}}]{The topic discusses social media debates on taxpayer-funded healthcare for \\ illegal immigrants, highlighting concerns about fairness, costs, and policies.}} \\
\midrule
\multirow{2}{*}{\makecell{}} & \multirow{2}{*}{\makecell{11-152 \\ (Topic 152)}} & \multirow{2}{*}{\makecell{}} & \makecell{Cost of Illegal Immigration}  & \makecell{59.85} & \makecell{14,342} \\
\cmidrule(lr){4-6}
&  &  & \multicolumn{3}{l}{\makecell[{{l}}]{This topic highlights concerns about the economic burden of illegal immigration, \\ taxpayer-funded benefits, and criticism of government immigration policies.}} \\
\midrule
\multirow{2}{*}{\makecell{}} & \multirow{2}{*}{\makecell{11-153 \\ (Topic 153)}} & \multirow{2}{*}{\makecell{}} & \makecell{U.S. Immigration and Funding Criticism}  & \makecell{61.48} & \makecell{14,601} \\
\cmidrule(lr){4-6}
&  &  & \multicolumn{3}{l}{\makecell[{{l}}]{Criticism of U.S. immigration policies focuses on national security, economic \\ burden, and fund allocation prioritizing immigrants over domestic needs.}} \\
\midrule
\multirow{2}{*}{\makecell{}} & \multirow{2}{*}{\makecell{11-154 \\ (Topic 154)}} & \multirow{2}{*}{\makecell{}} & \makecell{U.S. Immigration and Economic Concerns}  & \makecell{63.35} & \makecell{6,854} \\
\cmidrule(lr){4-6}
&  &  & \multicolumn{3}{l}{\makecell[{{l}}]{The topic critiques U.S. immigration policies under Biden, linking illegal \\ immigration to economic issues, crime, and social challenges.}} \\
\midrule
\multirow{2}{*}{\makecell{}} & \multirow{2}{*}{\makecell{11-155 \\ (Topic 155)}} & \multirow{2}{*}{\makecell{}} & \makecell{Illegal Immigration Debate}  & \makecell{56.78} & \makecell{1,552} \\
\cmidrule(lr){4-6}
&  &  & \multicolumn{3}{l}{\makecell[{{l}}]{The topic focuses on negative perceptions of illegal immigration in the U.S., \\ highlighting conservative, anti-immigrant views and related societal concerns.}} \\
\midrule
\multirow{2}{*}{\makecell{}} & \multirow{2}{*}{\makecell{11-156 \\ (Topic 156)}} & \multirow{2}{*}{\makecell{}} & \makecell{Unconstitutional Immigration Actions}  & \makecell{49.43} & \makecell{1,283} \\
\cmidrule(lr){4-6}
&  &  & \multicolumn{3}{l}{\makecell[{{l}}]{This topic discusses U.S. immigration issues, highlighting perceived \\ unconstitutional actions, policy criticism, and calls for legal adherence and \\ reform.}} \\
\midrule
\multirow{2}{*}{\makecell{}} & \multirow{2}{*}{\makecell{11-157 \\ (Topic 157)}} & \multirow{2}{*}{\makecell{}} & \makecell{Calls for Immigration Illegality}  & \makecell{50.57} & \makecell{1,450} \\
\cmidrule(lr){4-6}
&  &  & \multicolumn{3}{l}{\makecell[{{l}}]{The topic focuses on tweets advocating stricter U.S. immigration laws, \\ emphasizing illegal actions, stricter enforcement, and frustration with current \\ policies.}} \\

\bottomrule
\end{tabular}
\caption{Detailed Topic Discovery Results 14}
\label{table}
\end{table*}

\section{Appendix D: Interpretable Weekly Topic Trajectories}

\begin{table*}[t]
\centering
\renewcommand{\arraystretch}{1.2}
\begin{tabular}{>{\centering\arraybackslash}m{1.6cm}|>{\centering\arraybackslash}m{7.7cm}|>{\centering\arraybackslash}m{7cm}}
\noalign{\hrule height 1pt}
\textbf{Week} & \textbf{Increasing Toxicity Group} & \textbf{Reference Group} \\ \noalign{\hrule height 1pt}
4/17–4/23 & 41. DeSantis Immigration Policy Debate (42.36) & 148. Anti-Immigration Sentiment and Treason Accusations (71.79) \\ \hline
4/24–4/30 & 23. U.S. Immigration Politics and Controversy (71.11) & 54. Anti-Immigration Sentiment (66.81) \\ \hline
5/1–5/7 & 29. Supreme Court Immigration Rulings (31.79) & 149. Stopping Illegal US Immigration (50.79) \\ \hline
5/8–5/14 & 41. DeSantis Immigration Policy Debate (42.36) & 127. Anti-Immigration Sentiment (70.58) \\ \hline
5/15–5/21 & \multirow{3}{7.7cm}{\centering 22. Temporary Protected Status for Immigrants (30.18)} & 112. E-Verify and Immigration Reform (51.75) \\ \cline{1-1} \cline{3-3}
5/22–5/28 &  & 117. Hiring of Illegal Immigrants (56.68) \\ \cline{1-1}  \cline{3-3}
5/29–6/4 &  & 34. Amnesty for Illegal Immigrants (52.52) \\ \hline
6/5–6/11 & \multirow{2}{7.7cm}{\centering 96. Legal vs Illegal Immigration (47.28)} & 35. U.S. Immigration Policy Debate (53.13) \\ \cline{1-1}  \cline{3-3}
6/12–6/18 &  & \multirow{2}{7cm}{\centering 96. Legal vs Illegal Immigration (47.28)}  \\ \cline{1-1} \cline{2-2} 
6/19–6/25 &  6. Civics Test for Voting (50.33) & \\ \hline
6/26–7/2 & 14. Birthright Citizenship Debate (57.51) & 127. Anti-Immigration Sentiment (70.58) \\ \hline
7/3–7/9 & 23. U.S. Immigration Politics and Controversy (71.11) & 101. Immigration and Social Justice Issues (69.61) \\ \hline
7/10–7/16 & 41. DeSantis Immigration Policy Debate (42.36) & 77. Racist Immigration Rhetoric (67.84) \\ \hline
7/17–7/23 &  24. U.S. Immigration Policy Debate (40.33)  & \multirow{2}{7cm}{\centering 14. Birthright Citizenship Debate (57.51)}  \\ \cline{1-1} \cline{2-2} 
7/24–7/30 & 93. US Immigration and Border Crisis (38.42) &  \\ \hline
7/31–8/6 & \multirow{4}{7.7cm}{\centering 39. Secret Flights of Illegal Immigrants (56.67)}  & 155. Illegal Immigration Debate (56.78) \\ \cline{1-1}  \cline{3-3}
8/7–8/13 &  & 147. U.S. Immigration Concerns (61.67) \\ \cline{1-1}  \cline{3-3}
8/14–8/20 &  & \multirow{3}{7cm}{\centering 127. Anti-Immigration Sentiment (70.58)} \\ \cline{1-1}
8/21–8/27 &  &  \\ \cline{1-1} \cline{2-2}
8/28–9/3 & \multirow{5}{7.7cm}{\centering 147. U.S. Immigration Concerns (61.67)} &  \\ \cline{1-1}  \cline{3-3}
9/4–9/10 &  & 155. Illegal Immigration Debate (56.78) \\ \cline{1-1}  \cline{3-3}
9/11–9/17 &  & \multirow{4}{7cm}{\centering 149. Stopping Illegal US Immigration (50.79)} \\ \cline{1-1}
9/18–9/24 &  &  \\ \cline{1-1}
9/25–10/1 &  &  \\ \cline{1-2}
10/2–10/8 & 119. U.S. Border Immigration Concerns (65.13) &  \\ \hline
10/9–10/15 & 138. Immigration and National Security (68.13) & \multirow{2}{7cm}{\centering 127. Anti-Immigration Sentiment (70.58)} \\ \cline{1-1} \cline{2-2}
10/16–10/22 & 43. Immigration and Military Service (60.51) &  \\ \noalign{\hrule height 1pt}
\end{tabular}
\caption{Weekly Topic Trajectories – Increasing Toxicity vs. Reference Group (Topic Toxicity Scores)}
\label{tab:qual_inc}
\end{table*}

\begin{table*}[t]
\centering
\renewcommand{\arraystretch}{1.2}
\begin{tabular}{>{\centering\arraybackslash}m{1.6cm}|>{\centering\arraybackslash}m{7.05cm}|>{\centering\arraybackslash}m{7.65cm}}
\noalign{\hrule height 1pt}
\textbf{Week} & \textbf{Decreasing Toxicity Group} & \textbf{Reference Group} \\ \noalign{\hrule height 1pt}
4/17–4/23 & 35. U.S. Immigration Policy Debate (53.13) & 50. U.S. Border Wall Debate (54.32) \\ \hline
4/24–4/30 & 67. Enforcement of Immigration Laws (49.11) & \multirow{2}{7.65cm}{\centering 88. Republican Rhetoric on Immigration (52.67)} \\ \cline{1-1} \cline{2-2}
5/1–5/7 & 5. Ilhan Omar Immigration Controversy (73.93) &  \\ \hline
5/8–5/14 & 96. Legal vs Illegal Immigration (47.28) & \multirow{2}{7.65cm}{\centering 24. U.S. Immigration Policy Debate (40.33)} \\ \cline{1-1} \cline{2-2}
5/15–5/21 & 143. U.S. Immigration Border Crossings (47.83) &  \\ \hline
5/22–5/28 & 34. Amnesty for Illegal Immigrants (52.52) & \multirow{2}{7.65cm}{\centering 22. Temporary Protected Status for Immigrants (30.18)} \\ \cline{1-1} \cline{2-2}
5/29–6/4 & \multirow{2}{7.05cm}{\centering 96. Legal vs Illegal Immigration (47.28)} &  \\ \cline{1-1} \cline{3-3}
6/5–6/11 &  & 41. DeSantis Immigration Policy Debate (42.36) \\ \hline
6/12–6/18 & \multirow{3}{7.05cm}{14. Birthright Citizenship Debate (57.51)} & 88. Republican Rhetoric on Immigration (52.67) \\ \cline{1-1}  \cline{3-3}
6/19–6/25 &  & \multirow{2}{7.65cm}{\centering 23. U.S. Immigration Politics and Controversy (71.11)} \\ \cline{1-1} 
6/26–7/2 &  &  \\ \hline
7/3–7/9 & 37. Discrimination and Immigration Laws (52.73) & 84. Merit-Based Immigration Debate (43.18) \\ \hline
7/10–7/16 & 15. Ted Cruz Immigration Hypocrisy (62.98) & 115. Labor Shortage and Immigration (44.68) \\ \hline
7/17–7/23 & 14. Birthright Citizenship Debate (57.51) & \multirow{2}{7.65cm}{\centering 93. US Immigration and Border Crisis (38.42)} \\ \cline{1-1} \cline{2-2}
7/24–7/30 & 67. Enforcement of Immigration Laws (49.11) &  \\ \hline
7/31–8/6 & 96. Legal vs Illegal Immigration (47.28) & 24. U.S. Immigration Policy Debate (40.33) \\ \hline
8/7–8/13 & 76. Asylum Seekers and Immigration (47.88) & 70. DeSantis' Migrant Relocation Controversy (52.74) \\ \hline
8/14–8/20 & 72. Immigration Detention Criticism (39.79) & 34. Amnesty for Illegal Immigrants (52.52) \\ \hline
8/21–8/27 & \multirow{2}{7.05cm}{\centering 23. U.S. Immigration Politics and Controversy (71.11)} & 33. Reagan's Immigration Amnesty Policy (50.24) \\ \cline{1-1} \cline{3-3}
8/28–9/3 &  & 41. DeSantis Immigration Policy Debate (42.36) \\ \hline
9/4–9/10 & 96. Legal vs Illegal Immigration (47.28) & 68. US-Mexico Border Crisis (39.26) \\ \hline
9/11–9/17 & 58. U.S. Immigration Court Backlog (38.2) & 143. U.S. Immigration Border Crossings (47.83) \\ \hline
9/18–9/24 & 29. Supreme Court Immigration Rulings (31.79) & 33. Reagan's Immigration Amnesty Policy (50.24) \\ \hline
9/25–10/1 & 35. U.S. Immigration Policy Debate (53.13) & 23. U.S. Immigration Politics and Controversy (71.11) \\ \hline
10/2–10/8 & 29. Supreme Court Immigration Rulings (31.79) & 88. Republican Rhetoric on Immigration (52.67) \\ \hline
10/9–10/15 & 72. Immigration Detention Criticism (39.79) & 98. Western Intervention and Migration (58.25) \\ \hline
10/16–10/22 & 58. U.S. Immigration Court Backlog (38.2) & 101. Immigration and Social Justice Issues (69.61) \\ \noalign{\hrule height 1pt}
\end{tabular}
\caption{Weekly Topic Trajectories – Decreasing Toxicity vs. Reference Group (Topic Toxicity Scores)}
\label{tab:qual_dec}
\end{table*}

\end{document}